\documentclass[camera,letterpaper,nomarginnotes,nonarrowgutter]{jpaper}
\usepackage{color,soul}
\usepackage{amsmath,amssymb,amsfonts}
\usepackage{algorithmic}
\usepackage{graphicx}
\usepackage{textcomp}
\usepackage{xcolor}
\usepackage{fancyhdr}
\usepackage{xspace}
\usepackage[colorlinks = true,
            linkcolor = blue,
            urlcolor  = blue,
            citecolor = blue,
            anchorcolor = blue]{hyperref}
\definecolor{amber}{rgb}{1.0, 0.49, 0.0}
\definecolor{awesome}{rgb}{1.0, 0.13, 0.32}
\definecolor{dollarbill}{rgb}{0.52,0.73,0.4}
\definecolor{moegi}{rgb}{0.357, 0.537, 0.188}
\definecolor{burgundy}{rgb}{0.5, 0.0, 0.13}
\definecolor{ballblue}{rgb}{0.13, 0.67, 0.8}
\definecolor{ups-truck}{rgb}{0.53, 0.28, 0.21}
\definecolor{airforceblue}{rgb}{0.36, 0.54, 0.66}
\definecolor{cadmiumgreen}{rgb}{0.0, 0.42, 0.24}
\definecolor{darkcyan}{rgb}{0.0, 0.55, 0.55}
\definecolor{caribbeangreen}{rgb}{0.0, 0.8, 0.6}
\definecolor{flamingopink}{rgb}{0.99, 0.56, 0.67}
\definecolor{jazzberryjam}{rgb}{0.65, 0.04, 0.37}
\definecolor{mediumpersianblue}{rgb}{0.0, 0.4, 0.65}
\definecolor{coolblack}{rgb}{0.0, 0.18, 0.39}
\definecolor{bleudefrance}{rgb}{0.19, 0.55, 0.91}

\newcommand{\agy}[1]{{#1}}
\newcommand{\agya}[1]{\textcolor{black}{#1}}

\newcommand{\jk}[1]{\textcolor{black}{#1}}
\newcommand{\jkt}[1]{\textcolor{black}{#1}}

\usepackage{cite}
\usepackage{multirow}
\usepackage{hhline}
\usepackage{tikz}

\usepackage[compact]{titlesec}
\usepackage{footnote}
\usepackage{xfrac}
\usepackage{balance}

\newcommand{\hy}[2][yellow]{{%
{#2}}%
}
\newcommand{\om}[1]{\textcolor{black}{#1}}
\newcommand{\oma}[1]{\textcolor{black}{#1}}
\newcommand{\js}[1]{\textcolor{black}{#1}}
\newcommand{\hj}[1]{\textcolor{black}{#1}}
\newcommand{\hja}[1]{\textcolor{black}{#1}}
\newcommand{\hjb}[1]{\textcolor{black}{#1}}

\newcommand{\tech}{DarkGates\xspace}

\title{\Large {\tech}: A Hybrid Power-\jk{G}ating Architecture to Mitigate \jk{the} \\ \jk{Performance Impact} of Dark-Silicon in High Performance Processors}

\author{\vspace{-15pt}\\
{Jawad Haj Yahya$^{1,4,}$\textsuperscript{\textsection}} \quad
{Jeremie S. Kim$^{1}$} \quad
{A. Giray Ya\u{g}l{\i}k\c{c}{\i}$^{1}$} \quad
{Jisung Park$^{1}$}\\
\vspace{-9pt}\\
{Efraim Rotem$^{2}$} \quad
{Yanos Sazeides$^{3}$} \quad
{Onur Mutlu$^{1}$}
\vspace{12pt}\\
{\fontsize{10}{11}\selectfont
$^{1}$\textit{ETH Zurich} \quad
$^{2}$\textit{Intel Corporation} \quad
$^{3}$\textit{University of Cyprus} \quad
$^{4}$\textit{Huawei Technologies - Zurich Research Center}
}
\vspace{-2pt}}

\begin{document}
\bstctlcite{IEEEexample:BSTcontrol}
\maketitle
\thispagestyle{plain}
\pagestyle{plain}
\begingroup\renewcommand\thefootnote{\textsection}
\footnotetext{The work was done when Jawad Haj-Yahya was at ETH Zurich.}
\endgroup

\begin{abstract}

To reduce the leakage power of inactive (dark) silicon components, modern processor systems shut-off these components' power supply using low-leakage transistors, called \emph{power-gates}.
Unfortunately, power-gates increase the system's power-delivery impedance and voltage guardband, limiting the system's maximum attainable voltage (i.e., $V_{max}$) and, \jk{thus}, the CPU core's maximum attainable frequency (i.e., $F_{max}$). As a result, systems that are performance constrained by the CPU frequency (i.e., $F_{max}$-constrained), such as high-end desktops, suffer significant performance loss due to power-gates.
%For example, a $100mV$ of additional voltage guardband due to power-gates can reduce the $F_{max}$ of a $95W$ TDP (thermal design power) system by more than $7\%$.\agycomment{I know I asked for this, but actually this example hurts abstract's flow. I think it is better to either make it shorter or completely remove it.}

To mitigate this performance loss, we propose \emph{\tech}, a hybrid system architecture that increases the performance of $F_{max}$-constrained systems while fulfilling their \jk{power} efficiency requirements. 
{\tech} is based on \emph{three} key techniques: i) bypassing on-chip power-gates using package-level resources \jk{(called \emph{bypass mode})}, ii) extending power management firmware to support operation either in bypass \jk{mode} or normal mode, and iii) introducing deeper idle power states.

We implement {\tech} on an Intel Skylake microprocessor for client devices and evaluate it using a wide variety of workloads. On a real 4-core Skylake system with integrated graphics, {\tech} improves the average performance of SPEC CPU2006 workloads across all thermal design power (TDP) levels ($35W$--$91W$)  between $4.2\%$ and $5.3\%$. %For example, {\tech} improves the performance of SPEC CPU2006 workloads by up to $8.1\%$ ($4.6\%$ on average) for a $91W$-TDP (the highest TDP) desktop. 
\tech maintains the performance of 3DMark workloads for desktop systems with TDP greater than $45W$ while for a $35W$-TDP (the lowest TDP) desktop it experiences only a $2\%$ degradation.    
In addition, {\tech} fulfills the requirements of the ENERGY STAR and the Intel Ready Mode energy efficiency benchmarks of desktop systems.

\end{abstract}

% \keywords{
% % \begin{sloppypar}
% Power Management; Power Gates; Voltage Guardband 
% % \end{sloppypar}
% }
% \begin{IEEEkeywords}
% component; formatting; style; styling;

% \end{IEEEkeywords}

% For peer review papers, you can put extra information on the cover
% page as needed:
% \ifCLASSOPTIONpeerreview
% \begin{center} \bfseries EDICS Category: 3-BBND \end{center}
% \fi
%
% For peerreview papers, this IEEEtran command inserts a page break and
% creates the second title. It will be ignored for other modes.
% \IEEEpeerreviewmaketitle

\section{Introduction}\label{sec:intro}

Due to the breakdown of Dennard scaling \cite{dennard1974design}, the \hjb{fraction} of a silicon chip that can operate at the \jk{maximum} attainable frequency \hjb{(within a fixed power limit)} \jk{reduces} with each process generation \cite{merritt2009arm,esmaeilzadeh2011dark}. As a result, processor architects need to ensure that, at any point in time, \jk{a} large fraction of \jk{a chip is} effectively \emph{dark} (i.e., idle) or dimmed (i.e., underclocked)\jk{,} which limits performance. \jk{To this end}, architects clock-gate idle components to eliminate their dynamic power \jk{consumption} or \emph{power-gate} components to reduce their leakage power \jk{consumption} and use the saved precious energy to power-up \jk{the necessary} resources or increase their frequency.

As opposed to clock-gating, power-gating has a significant effect on a processor's architecture. A power-gate is implemented using area-hungry low-leakage transistors that can \jk{shut off} the voltage supply to \jk{a target idle} circuitry. A power-gate's impedance should be as small as possible to reduce the voltage drop it causes when the target circuit is active\jk{, as the impedance} has a direct impact on the \jk{circuit's} supply voltage and power consumption. However, lowering the power-gate's impedance increases \jk{the power-gate's} area cost. The area of such power gates is non-trivial as it grows as a function of the \jk{circuit} area that is power-gated. For instance, the area of a low-impedance power-gate for a CPU core can lead to a significant increase (${>}5\%$) in the overall chip area~\cite{zelikson2011embedded, shockley1952unipolar,petrica2013flicker,ditomaso2017machine,rahman2006determination,flynn2007low}. Unfortunately, \jk{since} there is a limited area budget for placing power-gates, it is impractical to minimize a power-gate's impedance and, \jk{thus,} this impedance \jk{causes} voltage drops on the power-delivery network. To cope with that, designers increase the voltage guardband, \jk{which results} in increased power consumption when the system is active \cite{hu2004microarchitectural,heo2002dynamic}. This limits the maximum attainable voltage (i.e., $V_{max}$) and frequency (i.e., $F_{max}$) \cite{cho2016postsilicon,de2015fine}, which can result in considerable performance loss for systems that are constrained by the maximum \jk{attainable CPU core} frequency (i.e., \emph{$F_{max}$-constrained}), such as high-end desktops (e.g., \jkt{Intel} Skylake-S \cite{Skylake_D,bowman2009impact,lee2009optimizing}).

% To mitigate the performance loss in high performance processors due to power-gates
In this paper, we propose \emph{\tech}, a hybrid power-gating architecture to increase the performance of $F_{max}$-constrained systems while satisfying their \jk{power} efficiency requirements. 
{\tech} is based on \emph{three} key techniques. 
First, {\tech} bypasses the power-gates of $F_{max}$-constrained processors at the \emph{package} level by 
shorting gated and un-gated CPU \jk{core} power-delivery domains. This enables the sharing of 
1) the decoupling capacitors of the die (i.e., Metal Insulator Metal (MIM) \cite{2_burton2014fivr}) and \jk{the} package (i.e., decaps \cite{15_jakushokas2010power}), and 
2) the package routing resources among CPU cores, resulting in lower voltage drops, and improving voltage/frequency (i.e., V/F) curves.\footnote{Intel processors are individually calibrated in the factory to operate on a specific voltage/frequency and operating-condition curve specified for the individual processor \cite{intel_tdc}. Reducing the voltage guardband increases the effective voltage, which allows the processor to operate at higher frequency for the same voltage level \cite{cho2016postsilicon}.}
Second, {\tech} extends the power management firmware (\jk{e.g.}, Pcode~\cite{gough2015cpu}) algorithms to operate in \jk{two modes: 1) bypass mode,  which} increases the CPU \agy{cores'} voltage and frequency, and
2) \jk{normal mode, which utilizes} the power-gates to reduce leakage power of CPU cores.
Third, \tech \jk{enables} deeper system idle power \jk{states} (i.e., package C-states) to reduce energy consumption once the entire processor is idle. For example, we \jk{include} support for $C8$ package C-state \cite{intel_skl_dev,haj2018power} to allow turning off most of the processor's components once the entire processor is idle, which reduces the average power consumption of energy-efficiency benchmarks. 

We implement {\tech} on an Intel Skylake microprocessor\footnote{Intel Skylake \cite{anati2016inside} shares its microarchitecture with multiple processors in 2015-2020, such as Kaby Lake~\cite{kabylake_2020}, Coffee Lake~\cite{coffeelake_2020}, Cannon Lake~\cite{i38121u_cannonlake}.} for client devices and \jk{evaluate} it using a wide variety of SPEC CPU2006, graphics (3DMark), and energy efficiency workloads.
On a 4-core Skylake processor \jk{with integrated graphics engines,}
\tech{} improves the performance of SPEC CPU2006 workloads by up to $8.1\%$ ($4.6\%$ on average)\footnote{The performance gains of \tech are significant in highly-optimized systems like the Intel Skylake. 
It is important to note that all new microarchitectural optimizations (e.g., improvements in pipelining, branch prediction, and memory subsystem) in Skylake generated a $2.4\%$ average performance improvement~\cite{anandtech_ipc} over Broadwell (one generation older than Skylake) and $5.7\%$ over Haswell (two generations older than Skylake)~\cite{anandtech_ipc}.} for a $91W$ thermal design power (TDP) desktop system. 
\tech{} maintains the performance of 3DMark workloads for desktop systems with a TDP higher than $45W$\jk{. For} a $35W$ TDP (the lowest TDP) desktop, \tech{}  degrades performance by \jk{only} $2\%$.    
In addition, {\tech} fulfills the ENERGY STAR (energy efficiency standard \cite{energy_star,energy2014energy}), the  Intel Ready Mode Technology (RMT \cite{intel_rmt,bolla2016assessing}) \jkt{energy efficiency benchmarks'} requirements. 

This work makes the following \textbf{contributions}:
\begin{itemize}

\item To our knowledge, {\tech} is the first work \jkt{that provides} \jk{a} hybrid power-gating architecture to increase the performance of systems that are constrained by the maximum \jk{attainable} CPU core frequency (i.e., \emph{$F_{max}$-constrained}), such as high-end  desktops.

\item We present the implementation of {\tech} on \jkt{the} Intel Skylake microprocessor for client devices, showing the three key techniques required to realize \tech and their overhead.

\item We perform an experimental evaluation of {\tech} on a real \jk{4-core Intel Skylake system and} clearly establish \jkt{{\tech}'} performance and energy \jkt{benefits} over a baseline system without it.

\end{itemize}

\section{Background}\label{sec:background}
We provide brief background on the architecture, power delivery networks, and design limits in modern client processors such as Intel Skylake \cite{anati2016inside, haj2016fine}, Kaby Lake \cite{kabylake_2020}, Coffee Lake \cite{coffeelake_2020}, and Cannon Lake \cite{i38121u_cannonlake}. 

\subsection{Client Processor Architecture}\label{sec:client_arch}
\begin{figure}[ht]
\begin{center}
\includegraphics[trim=0.5cm 0.6cm 0.5cm 0.6cm, clip=true,width=0.9\linewidth, keepaspectratio]{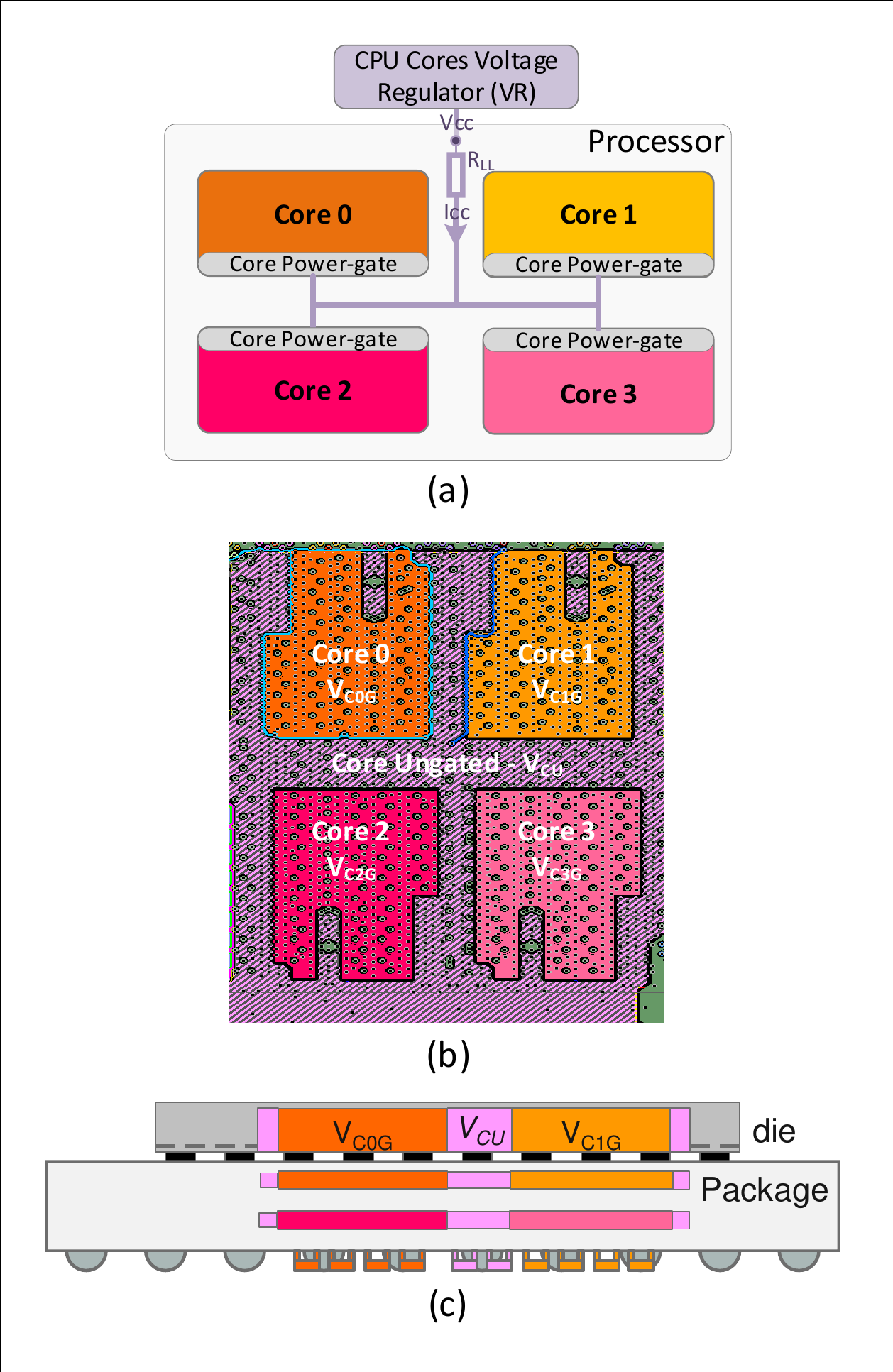}
\caption{(a) Architecture overview of recent Intel client processors. 
All cores share the same voltage regulator (VR). 
Each CPU core has a power-gate (PG) for the entire core. 
(b) Package layout showing an ungated main voltage domain ($V_{CU}$) arriving from the CPU core VR that feeds four power-gated voltage domains (one for each CPU core, $V_{C0G}$, $V_{C1G}$, $V_{C2G}$, and $V_{C3G}$). 
(c) Side view of die and package showing 1) the ungated main voltage domain ($V_{CU}$) and two cores' voltage domains ($V_{C0G}$ and $V_{C1G}$), and 2) the package's decoupling capacitors~\cite{15_jakushokas2010power}.}
\label{microarch}
\end{center}
\end{figure}

A high-end client processor is a system-on-chip (SoC) that typically integrates three main domains 
into a single chip: 1) compute (e.g., CPU
cores and graphics engines), 2) IO, and 3) memory {system}. 
Fig. \ref{microarch}(a) shows the architecture used in recent Intel processors (e.g., Skylake \cite{11_fayneh20164,yasin2019metric,anati2016inside}, Coffee Lake \cite{coffeelake_2020}, and Cannon Lake \cite{i38121u_cannonlake}) with a focus on CPU cores.

\noindent 
\textbf{Power Management.} 
The processor includes one central power management unit (PMU) and one local PMU per CPU core.
The central PMU is responsible for several power-management activities, such as dynamic voltage and frequency scaling (DVFS)~\cite{gonzalez1996energy,haj2018power,hajsysscale,rotem2011power}. 
The central PMU has several interfaces to on-chip and off-chip components, such as to 1) the motherboard voltage regulator (VR), i.e., serial voltage identification (SVID)~\cite{rotem2012power,rotem2015intel,haj2018power,gough2015cpu}, to control the voltage level of the VR, 2) the phase-locked-loop, to control the clock frequency, and 3) each core's local PMU, to communicate power management commands and status report. 
The local PMU inside the CPU core is responsible for core-specific power management, such as clock gating, power gating control, and thermal reporting.

\noindent
\textbf{Clocking.} 
A phase-locked loop (PLL) supplies the clock signal to all CPU cores.
All CPU cores have the same maximum clock frequency~\cite{8th_9th_gen_intel,icelake2020,perf_limit_reasons,haj2018energy}.\footnote{Typically, Intel client processors that use fully-integrated voltage regulator (FIVR) power delivery, including Haswell and Ice Lake processors, have the same clock frequency domain for all cores~\cite{6_kanter2013haswell,icelake2020}.} 

\noindent \textbf{Power Gating.}
Power gating is a circuit-level technique to significantly reduce the leakage power of an idle circuit~\cite{hu2004microarchitectural,haj2018power,gough2015cpu}. 
A power-gate is implemented using low-leakage transistors that shut off the input voltage to the target circuit and comes with a \emph{power-area tradeoff}. 
A power-gate needs to be large enough to help reduce 1) the power delivery impedance ($R$), 2) the voltage drop (i.e., $IR$ drop), and 3) the operating voltage and power consumption of the target circuit when it is active. 
However, a power-gate for a large circuit (e.g., a CPU core) can consume significant chip area~\cite{zelikson2011embedded,shockley1952unipolar,petrica2013flicker,ditomaso2017machine,rahman2006determination,flynn2007low}.
A power-gate may involve an additional trade-off between leakage power consumption reduction and performance loss due to the latency to ungate the circuit (i.e., to open the power gate). 
Typically, the wake-up latency from the power-gated state can take a handful to tens of cycles~\cite{kahng2013many,gough2015cpu}. 
However, to reduce the worst-case peak in-rush current~\cite{chadha2013architectural,usami2009design,agarwal2006power,abba2014improved} and voltage noise of power-delivery (e.g., $di$/$dt$ noise~\cite{larsson1997di,gough2015cpu,haj2018power}) when waking up a power-gate, the power-gate controller applies a \emph{staggered} wake-up technique~\cite{agarwal2006power} that takes tens of nanoseconds (typically, $10$--$20ns$)~\cite{kahng2013many,akl2009effective,kahng2012tap}.

\begin{sloppypar}
\noindent 
\textbf{Power Budget Management (PBM).}
To keep the system running below a thermal design power (TDP) limit, the SoC PMU employs a \emph{power budget management (PBM)} algorithm to dynamically distribute the total power budget to each SoC domain~\cite{lempel20112nd,rotem2013power,rotem2012power,21_doweck2017inside,ranganathan2006ensemble,rotem2011power,zhang2016maximizing,isci2006analysis,hajsysscale,ananthakrishnan2014dynamically,haj2020flexwatts}. 
This allows each domain to operate within its allocated power budget.
For instance, CPU cores and graphics engines in the compute domain share the same power budget. 
When a graphics-intensive workload is executed, the graphics engines consume most of the compute domain's power budget. 
\end{sloppypar}

To keep the power consumption of the compute domain within its allocated power budget, the PMU applies DVFS to 1)~reduce the CPU cores' power consumption and 2)~increase the graphics engines' performance~\cite{rotem2013power,rotem2012power,kim2008system,rotem2015intel}.

\begin{sloppypar}
\noindent 
\textbf{System Idle Power States (C-states).}
The Advanced Configuration and Power Interface (ACPI)\footnote{ACPI is an industry standard that is widely used for OS-directed configuration, power management, and thermal management of computing systems.} defines a processor's \emph{idle power states}, commonly called \emph{C-states}~\cite{acpi}.
C-states are defined in two primary levels: 1)~the component level, such as thread ($TCi$), core ($CCi$), and graphics ($RCi$) C-states, and 2)~the system level, known as \emph{package C-states} ($PCi$ or $Ci$)~\cite{intel_skl_dev,haj2018power}.
\end{sloppypar}

A package C-state defines an idle power state of the system (consisting of the processor, chipset, and external memory devices). 
A system enters a specific package C-state depending on each system component's idle power state (\emph{component C-state}). 
Various levels of package C-states exist to provide a range of power consumption levels with various techniques, such as clock gating at the uncore level or a nearly {complete} shutdown of the system. 
The ACPI standard includes recommendations on the C-states, but manufacturers are free to define their C-states and the corresponding system behavior at each C-state.
In this work, we focus on the package C-states of the Intel Skylake architecture \cite{intel_skl_dev}, but similar idle power state definitions exist in other architectures (e.g.,  AMD{\cite{amd_specs}} and ARM {\cite{qcom2018}}). 
Table~{\ref{tab:p-states}} shows all package C-states of the Intel Skylake architecture and the major conditions under which the power management unit (PMU) places the system into each package C-state (a similar table exists in the Intel manual{\cite{intel_skl_dev}}). 

\begin{table}[t]
%\vspace{-5pt}
\renewcommand{\arraystretch}{1.05}
%\caption{Description of package C-states on Intel Skylake architecture and the corresponding entry/exit latency  of each power state. }
\caption{Package C-states in \om{the} Intel Skylake mobile SoC.}
% \vspace{-5pt}
\label{tab:p-states}
\resizebox{\linewidth}{!}{%
\begin{tabular}{|l||l|}
\hline
\textbf{\begin{tabular}[c]{@{}l@{}}Package\\ C-state\end{tabular}} & \textbf{{Major conditions to enter the package C-state}}                                                                                                                                                                                                                                                                                \textbf{\begin{tabular}[c]{@{}l@{}} \\ \end{tabular}} \\ \hline \hline
C0                                                                 & One or more cores or graphics engine \textbf{executing instructions}                                                                                                                                                                                                                                                                                                 \\ \hline
C2                                                                 & \begin{tabular}[c]{@{}l@{}}All cores in \textbf{CC3} (clocks off) \textbf{or deeper} and graphics engine \\ in \textbf{RC6} (power-gated). DRAM is \bf{active}.\end{tabular}                                                                                                                                                                                                         \\ \hline
C3                                                                 & \begin{tabular}[c]{@{}l@{}}All cores in \textbf{CC3 or deeper} and graphics engine in \textbf{RC6}. \\ Last-Level-Cache (LLC) may be flushed and turned off,\\ DRAM in \textbf{self-refresh}, most IO and memory \\ domain clocks are gated, some IPs and IOs can be \textbf{active} \\ (e.g., \textbf{DC and Display IO}).\end{tabular}                                                         \\ \hline
C6                                                                 & \begin{tabular}[c]{@{}l@{}}All cores in \textbf{CC6} (power-gated) or deeper and graphics \\ engine in \textbf{RC6}. LLC may be flushed and turned off, DRAM \\ in \textbf{self-refresh}, IO and memory domain clocks generators are \\ turned off. Some IPs and IOs can be  active \\ \textbf{(e.g., video decoder (VD) and display controller (DC))}.\end{tabular}                                                                                  \\ \hline
C7                                                                 & \begin{tabular}[c]{@{}l@{}}Same as Package C6 while some of the IO and memory \\ domain voltages are \textbf{power-gated}. \textbf{CPU core VR is ON}.\end{tabular}                                                                                                                                                                                                                                \\ \hline
C8                                                                 & \begin{tabular}[c]{@{}l@{}}Same as Package C7 with additional \textbf{power-gating} in the IO \\ and memory domains. \textbf{CPU core VR is OFF}. \end{tabular}                                                                                                                                                                                                                                    \\ \hline
C9                                                                 & \begin{tabular}[c]{@{}l@{}}Same as Package C8 while all IPs must be off. Most voltage \\ regulators' voltages are reduced. \\ \textbf{The display panel can be in panel self-refresh (PSR)~\cite{psr, haj-micro-2021}}.\end{tabular}                                                                                                                                                                                       \\ \hline
C10                                                                & \begin{tabular}[c]{@{}l@{}}\textbf{Same as Package C9} while \textbf{all SoC VRs} (except state \\ always-on VR) \textbf{are off}. \textbf{The display panel is off}.\end{tabular}                                                                                                                                                                                                               \\ \hline
\end{tabular}%
}
% \vspace{-10pt}
\end{table}

Typical \emph{desktop} systems based on processors prior to Intel Skylake (e.g., Haswell \cite{bolla2016assessing,hammarlund2014haswell} or Broadwell \cite{mosalikanti2015low}) support up to \emph{package $C7$}, while mobile systems (e.g., Haswell-ULT \cite{kurd2014haswell} or Broadwell-ULX \cite{deval2015power}) support up to $C10$.   

\subsection{Client Processor Packages and Die Sharing}

Architects of modern  client processors typically build a \emph{single} CPU core (with a built-in power-gate) architecture\footnote{
An Intel CPU core has nearly the same microarchitecture for client and server processors. 
Intel CPU core design is a single development project, leading to a master superset core. 
Each project has two derivatives, one for server and one for client processors~\cite{Skylake_die_server}.} that supports \emph{all} dies of a client processor family, and some of the dies are used to build different processor packages targeting different segments. 
For example, the Intel Skylake processor for high-end \emph{mobile} (i.e., Skylake-H \cite{Skylake_H}) and high-end \emph{desktop} (i.e., Skylake-S \cite{Skylake_D}) processors uses a single processor die~\cite{skl_dies,tam2018skylake,21_doweck2017inside} for all TDP ranges (from  $35W$ \cite{intel_skl__3_5} to $91W$ \cite{intel_skl_91}). 
Recent AMD  client processors use a similar strategy \cite{singh20173,singh2018zen,burd2019zeppelin,beck2018zeppelin,amd_zen2_10,amd_zen2_54}.
This design reuse is adopted for two major reasons.  
First, doing so allows system manufacturers to configure a processor for a specific segment using two main methods: 1) by configuring processor's TDP (known as configurable TDP \cite{cTDP,cTDP2,jahagirdar2012power} or cTDP) to enable the processor to operate at higher or lower performance levels, depending on the available cooling capacity and desired power consumption of the system and 2) by integrating one or more dies (e.g., CPU dies, chipset, and embedded-DRAM) into a single package that is optimized for a specific market segment. 
For example, a land grid array (LGA \cite{kujala2002transition}) package is used for desktops while a ball grid array (BGA \cite{guenin1995analysis}) package is used for laptops.     
Second, it reduces non-recurring engineering (NRE \cite{magarshack2003system}) cost and design complexity to  allow competitive product prices and enable the meeting of strict time-to-market requirements. 

\subsection{Power Delivery {Network} (PDN)} 
There are three commonly-used PDNs in recent high-end client processors~\cite{haj2020flexwatts,haj2021ichannels}: motherboard voltage regulators (MBVR)~\cite{rotem2011power,10_jahagirdar2012power,11_fayneh20164,12_howse2015tick}, integrated voltage regulators (IVR)~\cite{singh20173,singh2018zen,burd2019zeppelin,beck2018zeppelin,toprak20145,sinkar2013low}, and low dropout voltage regulators (LDO)~\cite{2_burton2014fivr,5_nalamalpu2015broadwell,tam2018skylake,icelake2020}. 
We describe aspects of the MBVR PDN here due to its simplicity. 
As shown in Fig.~\ref{microarch}(a), the MBVR PDN of a high-end client processor includes 1)~one motherboard voltage regulator (VR) for all CPU cores, 2)~a \emph{load-line} impedance ($R_{LL}$), and  3)~power-gates for each individual core.\footnote{
Fine-grained power-gates exist in a CPU core.
For example, a power-gate is implemented in each AVX unit (e.g., AVX512~\cite{mandelblat2015technology,intel_avx512}) inside a CPU core.} 
All CPU cores share the same VR~\cite{rotem2011power,10_jahagirdar2012power,11_fayneh20164,12_howse2015tick,haj2018energy}. 
For more details on state-of-the-art PDNs, we refer the reader to our recent prior work~\cite{haj2020flexwatts}.

\begin{figure}[ht]
\begin{center}
\includegraphics[trim=0.8cm 0.99cm 0.8cm 0.8cm, clip=true,width=1\linewidth,keepaspectratio]{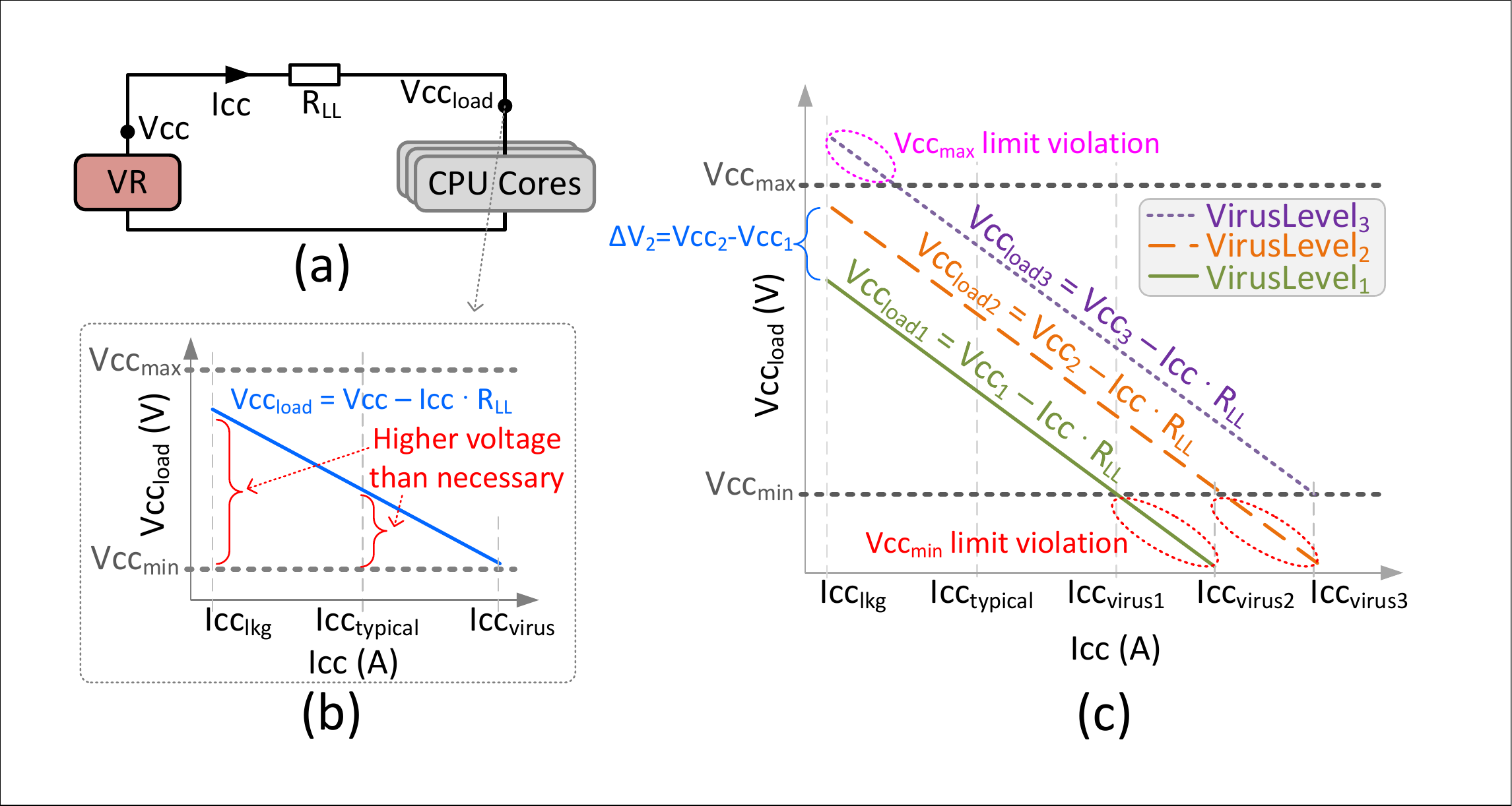}\\
\caption{Adaptive voltage guardband on modern processors. 
(a)~Simplified Power Delivery Network (PDN) model with a load-line. 
(b)~Voltage at the load is defined as: $Vcc_{load}=Vcc-R_{LL}\times{}Icc$, where $Vcc$ and $Icc$ are the voltage and current at the VR output, respectively.
(c)~Multi-level load-line with three power-virus levels. 
The voltage guardband is adjusted based on the power-virus level corresponding to the system state of the processor (e.g., number of active cores and instructions' computational intensity).}
\label{loadline}
\end{center}
% \vspace{-15pt}
\end{figure}

\noindent 
\textbf{Load-line.} 
Load-line or \emph{adaptive voltage positioning}~\cite{14_module2009and,intel_avp_2009,sun2006novel,tsai2015switching} is a model that describes the voltage and current relationship\footnote{In this model, short current bursts {result} in voltage droops~\cite{cho2016postsilicon,reddi2009voltage,reddi2010voltage}, which are filtered out by the decoupling capacitors~\cite{peterchev2006load}, while long current bursts are detected by the motherboard VR.} under a given system impedance, denoted by $R_{LL}$. 
Fig.~\ref{loadline}(a) describes a simplified power delivery network (PDN) model with a voltage regulator (VR), load-line ($R_{LL}$), and load (CPU Cores). 
$R_{LL}$ is typically {$1.6m\Omega$--$2.4m\Omega$} for recent client processors~\cite{8th_9th_gen_intel,haj2020flexwatts}. 
The voltage at the load is defined as $Vcc_{load}=Vcc-R_{LL}\times{}Icc$, where $Vcc$ and $Icc$ are the voltage and current at the VR output, respectively, as shown in Fig.~\ref{loadline}(b). 
From this equation, we can {observe} that the voltage at the load input ($Vcc_{load}$) decreases when the load's current ($Icc$) increases.
Due to this phenomenon, the PMU increases the input voltage ($Vcc$), i.e., {applies} a \emph{voltage guardband}, to a level that keeps the voltage at the load ($Vcc_{load}$) above the minimum functional voltage (i.e., $Vcc_{min}$) under even the most intensive load (i.e., when all active cores {are running} a workload that exercises the highest possible dynamic capacitance ($C_{dyn}$)). 
This workload is known as a \emph{power-virus}~\cite{haj2015compiler,fetzer2015managing,haj2020flexwatts} and results in the maximum possible current ($Icc_{virus}$). 
A typical application consumes a lower current $Icc_{typical}$ than $Icc_{virus}$. 
The minimal current that the processor can consume is the leakage current ($Icc_{lkg}$) once the clocks are gated (while the supply voltage is \emph{not} power-gated). 
In all cases where the current is lower than $Icc_{virus}$, the voltage drop (i.e., $R_{LL}\times{}Icc$) is smaller than when running a power-virus, which results in a higher load voltage $Vcc_{load}$ than necessary (as shown in Fig. \ref{loadline}(b)), leading to a power loss that increases quadratically with the voltage level. 

\begin{sloppypar}
\noindent \textbf{Adaptive Voltage Guardband, Icc$_{max}$, and Vcc$_{max}$.} To reduce the power loss {resulting} from a high voltage guardband when \emph{not} running a power-virus, due to the load-line effect, modern processors define \emph{multiple levels} of power-viruses depending on the maximum dynamic capacitance ($C_{dyn}$) that a system state (e.g., number of active cores and the computational intensity of running instructions) can draw. 
For each power-virus level, the processor applies a \emph{different} voltage guardband.  
%Power delivery network (PDN) of a processor handles short current conjunctions by the filter capacitor network on die, package and board \cite{zu2015adaptive}. For a high current event to be observed by the board and VR, it needs to last hundreds of nanoseconds to a few microseconds (few hundreds to a few thousands of core clock cycles), depending on PDN design. With this observation, the VR and its connection to the processor is shown in the simplified model of Fig. 3. 
%Fig. \ref{loadline}(a) describes a simplified power delivery network (PDN) model with load-line. The load-line or Adaptive Voltage Positioning (AVP) \cite{intel_avp_2009,sun2006novel,tsai2015switching} describes the behavior of the load voltage ($Vcc_{load}$) as it appears to the VR and motherboard\footnote{In this model, short current bursts results in voltage droops which are filtered out by the decoupling capacitors, while long current burst are observed by the motherboard VR}. $Vcc_{load}$ is the supply 
Fig. \ref{loadline}{(c)} illustrates the load-line model behavior of a processor with three power-virus levels denoted by $VirusLevel_1$, $VirusLevel_2$, and  $VirusLevel_3$ (where $VirusLevel_1 < VirusLevel_2 < VirusLevel_3$). The three power-virus levels represent multiple scenarios. For example, $VirusLevel_1$, $VirusLevel_2$, and $VirusLevel_3$ can represent one, two, and four active cores, respectively, for a processor with four cores. 
When the processor moves from one power-virus level to a higher/lower level, the processor increases/decreases the voltage by a voltage guardband ($\Delta V$). For example, when moving from $VirusLevel_1$ to  $VirusLevel_2$, the processor increases the voltage by $\Delta V_2$ as shown ({in} blue {text}) in Fig. \ref{loadline}{(c)}.
\end{sloppypar}

\subsection{Processor Design Limits}
\subsubsection{Thermal Limits}\label{sec:thermal_limits}
We describe the most important thermal limits that constrain the performance \js{of} modern processors. 

\noindent 
\textbf{Junction Temperature ($T_{jmax}$) Limit.}
As the processor dissipates power, the temperature of the silicon junction ($T_j$) increases. 
$T_j$ should be kept below the maximum junction temperature ($T_{jmax}$) as overheating may cause permanent damage to the processor. 
The processor measures the temperature and applies multiple techniques (e.g., PBM~\cite{lempel20112nd}, thermal throttling~\cite{haj2018power}, and catastrophic trip temperature~\cite{gough2015cpu}) to ensure that the temperature remains under the $T_{jmax}$ limit. 
In the worst case, the processor automatically shuts down when the silicon junction temperature reaches its operating limit~\cite{haj2018power,haj2018energy,rotem2015intel,intel_xtu_overclocking,intel_avp_2009}.

\noindent \textbf{Thermal Design Power (TDP).}
TDP (in watts) is the maximum sustainable power consumption under the maximum theoretical load (e.g., common applications, but not a power-virus) that the cooling solution of the system needs to be designed for~\cite{xie2014therminator,rotem2015power,isci2006analysis,haj2018power,cTDP,cTDP2}.

\subsubsection{Power Delivery Network (PDN) Limits}\label{sec:pdn_limits}
There are multiple PDN limits in a modern processor. We describe the most important ones. 

\noindent \textbf{VR Thermal Design Current (TDC).}
TDC is the the continuous load current, also known as maximum continuous current, thermal current, or second power limit (i.e., PL2 \cite{rotem2015intel}). TDC is the sustained current that the processor is capable of drawing indefinitely and defines the current to use for VR temperature assessment. In other words, TDC is the \js{maximum} amount of electrical current the VR must be able to supply while being thermally viable~\cite{intel_tdc,burd2019zeppelin,su2017high}.

\noindent \textbf{Power Supply and Battery Maximum Current Limit.}
The power supply unit (e.g., ATX power supply \cite{meisner2009powernap} or power brick \cite{rotem2015intel}) and/or device battery that supply current to the system VRs also have current limits. 
For example, the third power limit (i.e., PL3) is used for battery over-current protection \cite{rotem2015intel}.  

\noindent 
\textbf{VR Electrical Design Current (EDC) Limit.}
The power delivery of a modern processor is limited by EDC, also known as the maximum instantaneous current, peak current, $Icc_{max}$, or fourth power limit (i.e., PL4~\cite{rotem2015intel}). 
EDC is the maximum amount of current at any instantaneous short period of time that can be delivered by a motherboard VR or an integrated VR (e.g., FIVR~\cite{2_burton2014fivr}). 
EDC limit is typically imposed by the limited maximum current that the VRs can supply~\cite{gough2015cpu,2_burton2014fivr,haj2018power,skylakex,intel_avp_2009,naffziger2016integrated}.
Exceeding the EDC limit can result in irreversible damage to the VR \js{or} the processor chip, or tripping \js{the} VR's protection mechanism for excessive current that shut\js{s down} the system. 
Therefore, a combination of \emph{proactive} enforcement and platform design constraints must be used to prevent system failure~\cite{gough2015cpu,2_burton2014fivr,haj2018power,skylakex,intel_avp_2009,naffziger2016integrated,wright2006characterization,piguet2005low,zhang2014architecture,su2017high}.

\noindent 
\textbf{Maximum Current per Bump/Pin.} 
The amount of current that a processor's die/package can consume per voltage domain is limited by the maximum current a bump/pin can support~\cite{wright2006characterization,piguet2005low,zhang2014architecture}. 
For example, while integrated VRs mitigate the EDC limit by enabling a reduced input current for the processor \cite{haj2020flexwatts}, the maximum current of the processor can be limited by the maximum current for a bump/pin.

\noindent \textbf{Minimum Operating Voltage Limit ($V_{min}$).}
Operating below the $V_{min}$ limit can cause a processor to malfunction. Therefore, modern processors implement multiple techniques to prevent the voltage from dropping below  $Vcc_{min}$ due to, for example, $di$/$dt$ voltage fluctuations~\cite{joseph2003control,gough2015cpu,haj2018power,reddi2010voltage,reddi2009voltage, brooks2001dynamic,lefurgy2011active}.

\noindent 
\textbf{Maximum Operational Voltage Limit ($V_{max}$).}
Technology scaling has made modern integrated circuits more susceptible to reliability degradation phenomena such as Negative Bias Temperature Instability (NBTI), Electromigration (EM), and Time Dependent Dielectric Breakdown (TDDB) \cite{hu1996gate}. 
Degradation depends on many processes and environmental factors, but can be controlled by managing the circuit's temperature and voltage levels~\cite{mercati2013workload}. 
Processor manufacturers define a maximum operational voltage limit ($V_{max}$) that should not be exceeded to ensure the guaranteed processor lifespan and reliability.
For example, Intel allows exceeding $V_{max}$ when overclocking a system (e.g., via the BIOS \cite{intel_bios_overclocking} or the XTU tool~\cite{intel_xtu_overclocking}). 
This process is out of the processor's reliability specification and can shorten the processor's lifespan~\cite{intel_bios_overclocking,intel_xtu_overclocking,intel_avp_2009}.

\noindent 
\textbf{Voltage Droop Effect on Maximum Frequency ($F_{max}$).}
In an active CPU core, simultaneous operations in memory and/or logic circuits demand high current flow, which creates fast transient voltage droops from the nominal voltage ($V_{nom}$). 
The worst-case voltage droop can degrade the maximum attainable frequency at a given voltage since this requires additional voltage (droop) guardband ($V_{gb}$) above the nominal voltage to enable the CPU core to run at the target frequency. 
\js{If} the core voltage with the voltage guardband \js{becomes} higher than $V_{max}$ (i.e., $V_{nom} + V_{gb} > V_{max}$)\js{,} the power management unit reduces $F_{max}$, thereby reducing $V_{nom}$ to keep the CPU core voltage below $V_{max}$. Therefore, voltage guardband ($V_{gb}$) has a direct effect on the CPU core $F_{max}$.

\section{Motivation}
\label{sec:motiv}

We conduct experiments on two different \hj{system} setups to clearly motivate the productization of \tech{}. 

Our first setup is a real  Intel Broadwell processor \cite{5_nalamalpu2015broadwell}, the previous generation of our target Skylake processor \cite{21_doweck2017inside}. We configure the Broadwell processor to four Thermal Design Power (TDP) \hj{levels} and frequencies using post-silicon configuration tools
(see Sec. \ref{sec:method}).  In this experiment\hj{,} we reduce the voltage guardband of the CPU cores by $100mV$, allowing the power  budget  management algorithm (PBM, see Sec. \ref{sec:client_arch}) to increase the CPU cores' frequency for a given voltage while keeping the system power consumption below TDP and the voltage below the maximum \hj{operating} voltage limit ($V_{max}$). 
The goal of this experiment is to evaluate the potential performance benefits of increasing CPU core clock frequency \hj{as we increase} the effective voltage by reducing the voltage guardband. In this experiment\hj{,} we run the SPEC CPU2006 benchmarks\hj{, both} floating-point (fp) and integer (int) with base (single-core) and rate (all cores) modes \cite{SPEC2018}. 

Our second experimental setup is based on an in-house power delivery network simulator  (see Sec. \ref{sec:method}) that aims to evaluate the 
maximum possible reduction in system impedance when \hj{we bypass} the power-gates.

We make two key observations from these experiments:

\noindent \textbf{Observation 1.}
Reducing the voltage guardband (e.g., IR drop compensation) increases the effective voltage, which allows increasing the processor frequency with a negligible increase in power consumption. 

\begin{figure}[!h]
  \begin{center}
  \includegraphics[trim=0.8cm 0.8cm 0.8cm 0.8cm, clip=true,width=1.\linewidth,keepaspectratio]{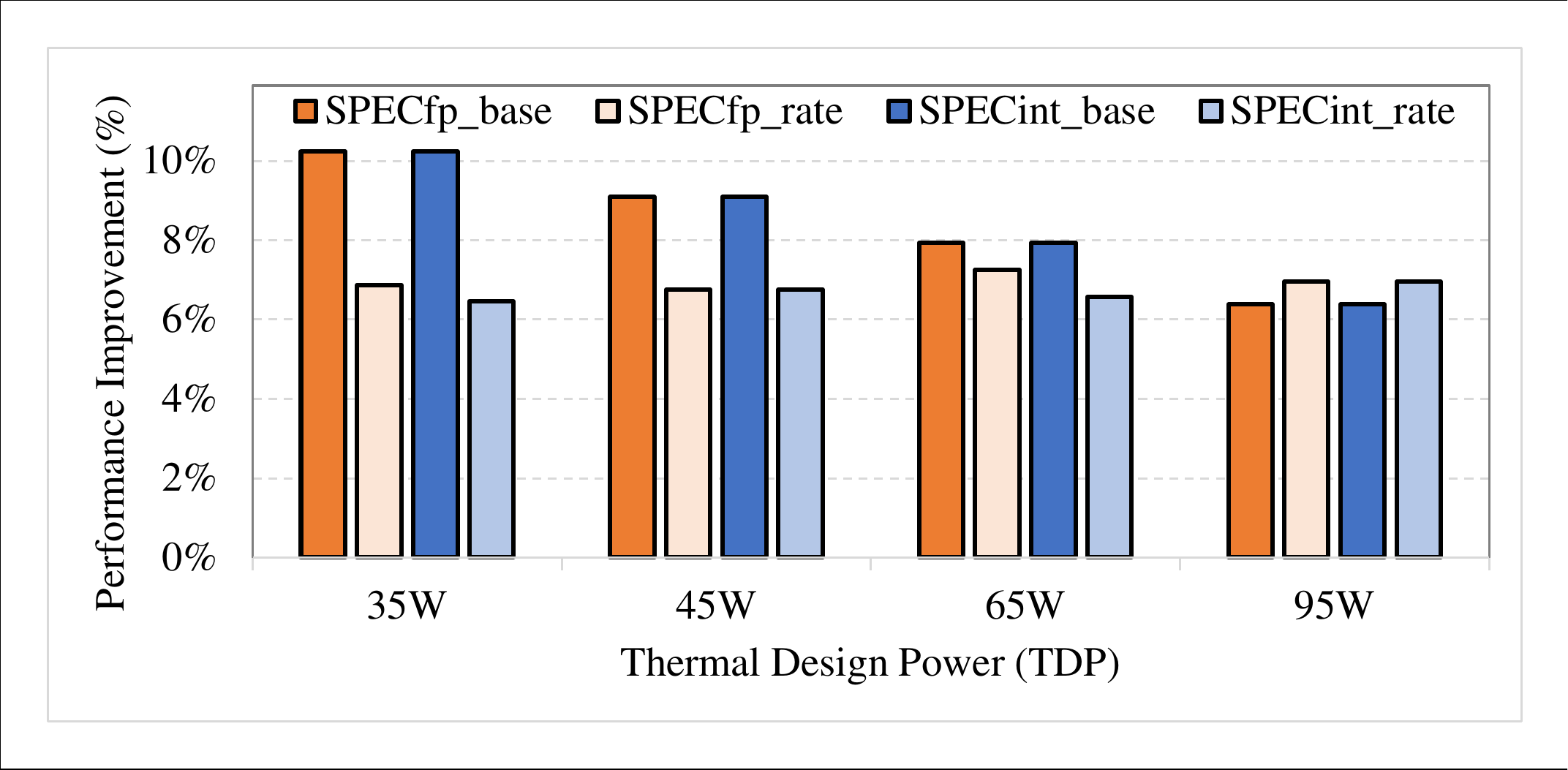}\\
  \caption{Average performance improvement of floating-point SPEC CPU2006 (floating-point (fp) and integer (int) \hj{benchmarks} with base \hj{(single-core)} and rate \hj{(all-cores)} modes) when \hj{we increase the} CPU core frequency\hj{, which is enabled} by increasing the effective voltage \hj{as a result of} reducing the voltage guardband by $100mV$.}\label{fig:motiv1}
  \end{center}
% \vspace{-15pt}
\end{figure}

Fig.~\ref{fig:motiv1} plots the performance impact of increasing \hj{the} CPU core frequency \hj{of an Intel Broadwell system}\agy{,} which is enabled by increasing the effective voltage as a result of reducing the voltage guardband by $100mV$.  We gather our results using SPEC CPU2006  benchmarks for four TDP levels.

We make five key observations from Fig. \ref{fig:motiv1}. 
First, the average performance of SPECfp and SPECint \hj{benchmarks} increases (by $6$--$10\%$) \hj{as} the frequency of the system \hj{increases for each given TDP level}. 
Second, the system can run at \hj{a} higher frequency with the same CPU core voltage level \emph{without} exceeding the TDP limit \hj{since the effective voltage increases once we reduce the voltage guardband}.
Third, the performance of high\hj{-}TDP (i.e., $95W$) configurations  increases \hj{even though} these systems are typically limited by  $V_{max}$ (i.e., $F_{max}$-constrained) \hj{since the effective $V_{max}$ voltage increases once we reduce the voltage guardband}.    
Fourth, the lower the TDP\hj{,} the higher the performance gain of the SPEC benchmarks in the base \hj{(i.e., single-core)} mode. \hj{This is because the relative increase of frequency in steps of $100MHz$ granularity until reaching the TDP limit is higher as the TDP (and baseline frequency) level is lower.}
Fifth, the high TDP (i.e., $95W$) performance gain of the SPEC benchmarks in the rate \hj{(i.e., all-cores)} mode is higher than \hj{that in} \agy{the} base mode. \hj{This is because we can increase the frequency of all cores to the maximum attainable frequency corresponding to the improved $V_{max}$ (due to the reduced the voltage guardband) without exceeding \hj{the} TDP level since these \hj{systems} are typically  $V_{max}$ limited.}   

We conclude that reducing the voltage guardband can significantly improve the performance of both thermally\hja{-}limited  systems (e.g., $35W$ \hja{TDP}) and $F_{max}$-constrained systems (e.g., $95W$ \hja{TDP})\hj{, based on experiments on real Intel Broadwell systems.}
%\agycomment{These conclusions do not come out naturally because fmax-constrained systems are not explicitly mentioned in the previous paragraph. Maybe, we need a bit of insights in the previous paragraph after each observation... I'm not sure.}

\begin{figure}[!ht]
  \begin{center}
  \includegraphics[trim=0.1cm 0.0cm 0.0cm 0.0cm, clip=true,width=.99\linewidth,keepaspectratio]{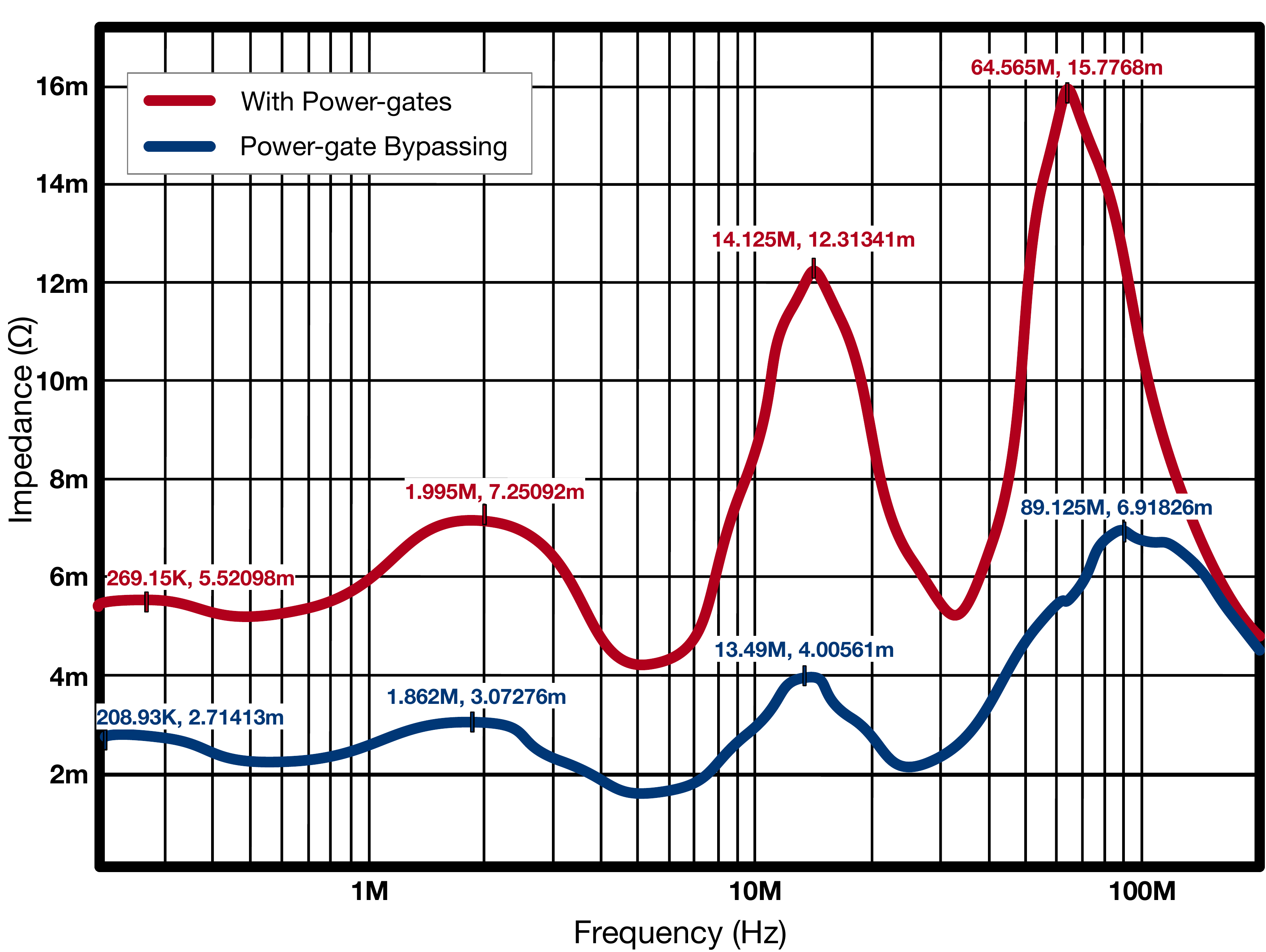}\\
  \caption{The impedance-frequency profile of a Skylake system that 1) uses  power-gates (red) and 2) bypasses the power-gates (blue). The system that uses the power-gates has approximately ${2}\times$ the impedance of a system that bypasses the power-gates.}\label{fig:motiv2}
  \end{center}
%  \vspace{-15pt}
 \end{figure}

\noindent \textbf{Observation 2.}  
While power-gating is an effective technique to reduce leakage power of idle CPU cores, we observe that power-gates can significantly increase system impedance, which increases voltage drop (e.g., resistive voltage drop, $IR$ drop), thereby requiring higher voltage guardband\hj{s} to compensate \hj{for the higher}  voltage drop. Fig.~\ref{fig:motiv2} shows the impedance-frequency profile \cite{mansfeld1981recording,leng2014gpuvolt,ketkar2009microarchitecture} of \hj{two simulated Intel} Skylake system\hj{s:} 1) \hja{one} that uses power-gates (red) and 2) \hja{another} that bypasses the power-gates (blue). The system that uses the power-gates has approximately $2\times$ the impedance of a system that bypasses the power-gates. Therefore, a system that uses the power-gates requires approximately \agy{$2\times$} the voltage guardband of a system that bypasses the power-gates. 

We conclude that bypassing the power-gates can reduce system impedance by approximately \agy{$2\times$}, which allows reducing the voltage drop guardband by approximately $2\times$. 

 \noindent\textbf{Summary.} Our experimental results clearly demonstrate that modern desktop systems face a significant challenge \hj{against} achieving their potential performance and TDP utilization due to the voltage drops on power gates\hj{. Even though} these \hj{voltage} drops are completely preventable (e.g., if power-gates are removed), \hj{they} still exist in real desktop processors. This \hj{is} because reusing the same processor die \hj{for a wide} variety of systems \hj{(e.g., from mobile to desktop to server)} with \hj{built-in} power-gates configuration is economically preferable. 
% Given these observations, we argue that a hybrid power management approach is needed to mitigate the inefficiencies in modern desktop processors while keeping the die reuse for its economical benefits. 
 
Based on our key observations, we conclude that a \textit{hybrid power\hj{-gating} approach} is necessary to mitigate the power-gating inefficiencies in current client processors. Our \textbf{goal} is to provide \hj{such an} approach that 1) reduces the voltage guardband overhead of power-gates in high\hj{-}performance \hj{systems} (e.g., \hj{high-end}  desktops), 
2) utilizes the reduced voltage guardband to increase the frequency of the CPU cores without increasing the baseline voltage or exceeding the TDP, and
3) provides low energy consumption for battery\hj{-}operated systems (e.g., laptops) and meets the energy-efficiency benchmarks\hja{'} requirements for desktop systems \hj{by reducing the leakage power consumption of idle cores}.

\section{{\tech} Architecture}
\label{sec:technique}

Based on our experimental \hj{analyses,} we propose \emph{\tech}, a hybrid system architecture that increases the performance of $F_{max}$-constrained systems while fulfilling their \hj{power} efficiency requirements. 

We design {\tech} with two design goals in mind: 
1) reduce CPU \hj{cores' power-delivery impedance, and, thus,} voltage drop\hj{,} to improve the V/F curve of high-end desktops, and
2) meet the energy efficiency requirements of desktop devices by enabling deeper package C-states. 

{\tech} achieves these two goals with \emph{three} key components.
The \emph{first component} of {\tech} is a \emph{Power-gates Bypassing} technique that effectively bypasses the power-gates of $F_{max}$-constrained processors at \hj{the} \emph{package} level by 
shorting gated and un-gated CPU \hj{core} power domains. This \hj{leads} to 1)  sharing of the decoupling capacitors of the die and package between CPU core, and 2)  sharing of package routing resources between CPU cores, resulting in lower voltage drops. \hj{The result is improved} voltage/frequency (i.e., V/F) curves.

The \emph{second component} of {\tech} is \hj{the improved} \emph{power management firmware} that is responsible for extending the power management algorithms to operate in \hj{two modes}: 1) bypass mode\hj{,} which increases the CPU core voltage and frequency by utilizing the improved V/F curves, and
2) normal mode\hj{,} which utilizes the power-gates to reduce \hj{the} leakage power of CPU cores.

The \emph{third component} of {\tech} is a \emph{new deep package C-state} for desktop systems that reduces  energy consumption once the entire processor is idle. This leads to improved average power consumption \hj{for} desktop energy-efficiency benchmarks.

The three components of \tech work together to increase  the  performance of  $F_{max}$-constrained systems  while  fulfilling  the energy efficiency requirements. We describe them in detail in the next three subsections.

\subsection{Power-gate Bypassing}\label{sec:pg_bypassing}

\begin{figure}[ht]
\begin{center}
\includegraphics[trim=0.5cm 0.6cm 0.5cm 0.6cm, clip=true,width=0.9\linewidth,keepaspectratio]{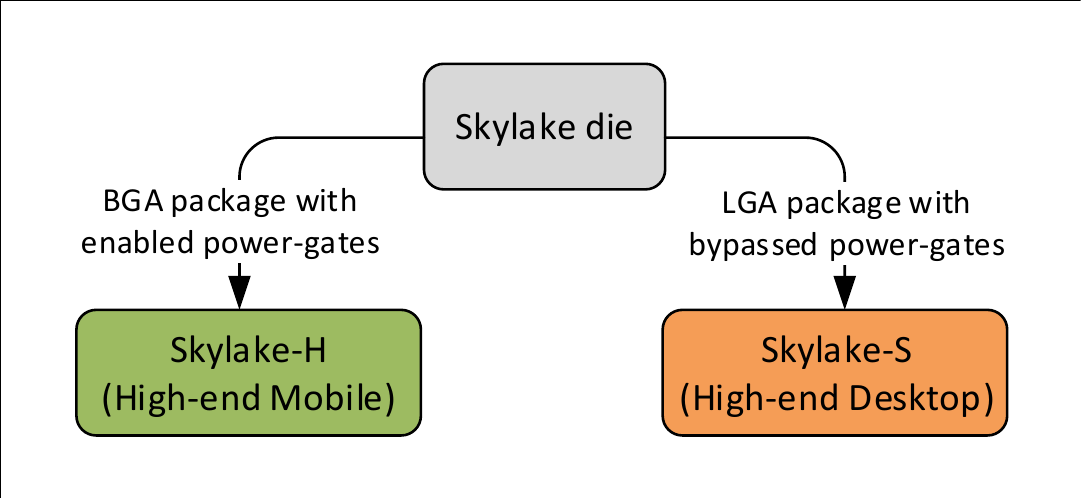}
% \vspace{-5pt}
\caption{The {\tech} hybrid power-gating architecture uses the same Skylake die to build 1) a dedicated package for Skylake-H (used for high-end mobile systems) with the power-gates enabled and 2) a dedicated package for Skylake-S (used for high-end desktop systems) that bypasses the power-gates.}\label{pg_bypassed_enabled}
\end{center}
% \vspace{-15pt}
\end{figure}

\hj{The} {\tech} Power-gate Bypassing technique is responsible for reducing CPU cores' voltage drop \hj{in}  $F_{max}$-constrained  systems (\hj{e.g.,} high-end desktops) by reducing system impedance. To do so, the technique uses the same \hj{Intel} Skylake die to build 1) a dedicated package for Skylake-H (used for high-end mobile systems) with the power-gates enabled and 2) a dedicated package for Skylake-S (used for high-end desktop systems) that bypasses the power-gates\hj{,} as \hj{shown} in Fig. \ref{pg_bypassed_enabled}. This architecture is feasible since client processors typically share the same die between multiple mobile and desktop products. Specifically, the same die is used for \hj{both} \hja{Intel} Skylake high-end mobile \hja{systems} (Skylake-H) and Skylake desktop \hja{systems} (Skylake-S).\footnote{Some of Skylake products even integrates \hj{an} additional embedded DRAM die into the same processor package \cite{21_doweck2017inside}.}

\begin{figure}[!ht]
\begin{center}
\includegraphics[trim=0.5cm 0.6cm 0.5cm 0.6cm, clip=true,width=0.9\linewidth,keepaspectratio]{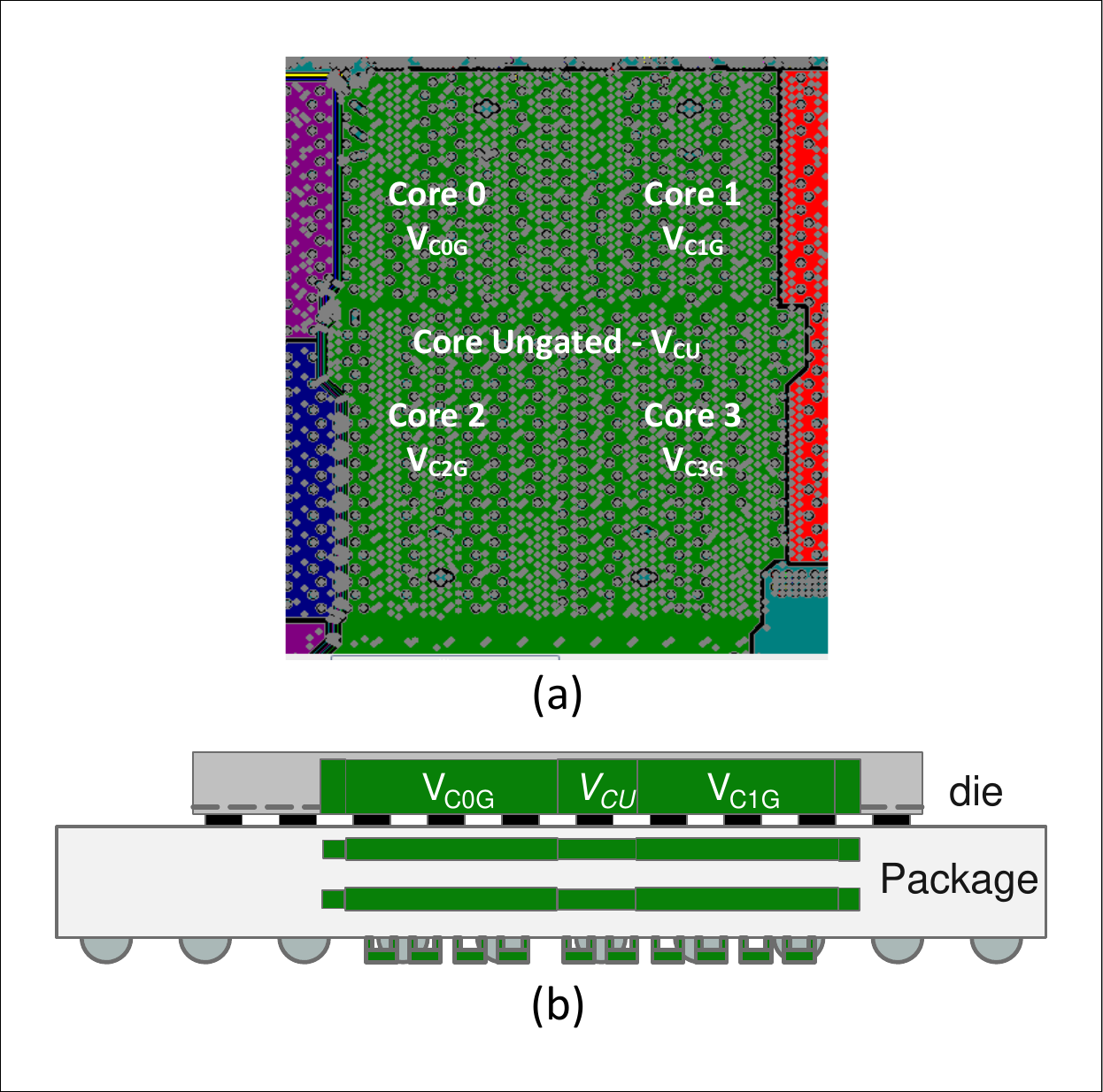}
%  \vspace{-10pt}
\caption{(a) Package layout of Skylake-S (high-end desktop) showing a single voltage domain for the cores (green), which combines the Core \hj{Ungated} voltage domain ($V_{CU}$) and each of CPU core voltage domains, $V_{C0G}$, $V_{C1G}$, $V_{C2G}$, $V_{C3G}$). (b) Side view of die and package showing combined voltage domain in die, package's substrate, and package's decoupling capacitors.}\label{pg_bypass}
\end{center}
% \vspace{-10pt}
\end{figure}

As \hj{shown} in Fig. \ref{pg_bypass}, the desktop package combines \hj{into a single voltage domain} the five \hj{voltages} used in the mobile package \hj{shown} in Fig. \ref{microarch} (i.e., the core ungated voltage domain ($V_{CU}$) and the per-core gated voltage domains, $V_{C0G}$, $V_{C1G}$, $V_{C2G}$, and $V_{C3G}$). 
To do so, the desktop package effectively shorts the \emph{four} gated CPU \hj{cores'} voltage domains and the ungated voltage domain into a single domain. 

The single voltage domain architecture leads to 1)  sharing of the decoupling capacitors of the die (i.e., Metal Insulator Metal (MIM) \cite{2_burton2014fivr})  and package (i.e., decaps \cite{15_jakushokas2010power}) between CPU \hj{cores}, and 2)  sharing of package routing resources between CPU cores. This architecture  results in reducing both resistive and inductive voltage drops \cite{zu2015adaptive,zou2018efficient,leng2015gpu,reddi2010voltage}. As discussed in Sec. \ref{sec:pdn_limits}, reducing the voltage drop reduces the voltage guardband, which improves the voltage/frequency (i.e., V/F) curves. Improved V/F curves lead to \hj{a higher} frequency \hj{(and, thus, performance)} for $F_{max}$-constrained systems.

\subsection{\hj{Improved} Power Management Algorithms}
\label{sec:pm_alg}
\tech architecture requires the adjustment of three main components inside the power management unit (PMU) firmware (i.e., Pcode \cite{gough2015cpu}).

\begin{sloppypar}
First, \tech requires the adjustment of DVFS firmware power management algorithms (e.g., P-state, Turbo) that enables the transition from one frequency/voltage operating point to another \cite{hajsysscale,haj2018energy}. Since Power-gate Bypassing \hj{(Sec. \ref{sec:pg_bypassing})} improves the V/F curves (i.e., for a \hj{given} voltage level\hj{,} \hja{it increases} the maximum attainable frequency), the DVFS algorithms should be adjusted to take into account the new V/F curves for desktop system\footnote{The firmware (i.e., Pcode \cite{gough2015cpu}) can recognize the target system (i.e., mobile vs. desktop) based on fuses available to the firmware.} by allowing \hj{the appropriate} higher frequency at \hj{any given} voltage level. In particular, these changes allow increasing the maximum attainable frequency (i.e., $F_{max}$) for \hj{systems} that are limited by the maximum reliable voltage (i.e., $V_{max}$). An additional advantage of the improved V/F curves is that a CPU core can run at a given frequency with \hj{a} lower voltage level, which reduces the processor power \hj{consumption} in the active state.
\end{sloppypar}

Second, \tech requires \hj{adjustments in} the power budget management algorithm (PBM) \cite{lempel20112nd,rotem2013power,rotem2012power,21_doweck2017inside,ranganathan2006ensemble,rotem2011power,zhang2016maximizing,isci2006analysis} \hj{(discussed in Sec. \ref{sec:background})}. 
%PBM distributes the power budget between CPU cores and the graphics engines according to the characteristics of the workloads that run on these units. 
\hj{In particular, in the \tech architecture, \hj{the} PBM algorithm needs to take into account the additional power consumption due to the leakage of inactive cores (i.e., ungated idle CPU \hja{core}), which can decrease the effective power budget that is allocated to the active cores and/or graphics engines. The reduced power budget can decrease the frequency of a thermally-limited compute domain}. 
%PBM is designed to keep the average power consumption of the processor within the allocated power budget (i.e., TDP). 

\hj{Third}, \tech can change the \hja{processor's} lifetime reliability \cite{song2014architectural,swaminathan2017bravo,kim2016learning}  due to \hj{the} additional power and temperature resulting from bypassing the power gates. \hj{On the} one hand, \tech's Power-gate Bypassing that combines cores' \hja{otherwise separated} voltage domains into a single voltage domain can improve the maximum current provided to each core since\hj{,} with \tech, all bumps are shared between the cores, which \hj{alleviates} the electromigration (EM) \hj{issues} \cite{kim2016learning}. 
On the other hand,  Power-gate Bypassing keeps cores powered-on in workloads \hj{where} one \hj{or} more cores are idle since these cores are normally power-gated in the baseline system (i.e., without \tech). Therefore, Power-gate Bypassing  1) increases the stress time of the CPU cores and 2) increases the junction temperature compared to baseline. As a result, \tech requires the adjustment of the reliability voltage guardband. Our reliability model shows that less than $5mV$/$20mV$ of \emph{additional reliability voltage guardband} is required to compensate \hj{for} the additional stress and temperature for $91W$/$35W$ (additional ${\sim}5^{\circ}C$), respectively.

\subsection{\hj{New} Package C-state \hj{for Desktops}}
As discussed in Sec. \ref{sec:client_arch}, \emph{desktop} systems in previous processor generations of Skylake support up to \hj{the} \emph{package $C7$} (defined in Table \ref{tab:p-states}). For example, the deepest package C-state that Haswell \cite{bolla2016assessing,hammarlund2014haswell} and Broadwell \cite{mosalikanti2015low}  desktop \hj{systems} support is package $C7$,  while Haswell \cite{kurd2014haswell} and Broadwell \cite{deval2015power} for mobile systems support up to $C10$. The difference in package C-state support between desktop and mobile is due to two major reasons. 
First, reducing energy consumption is critical for mobile \hj{systems} to meet battery life requirements for representative benchmarks (e.g., video playback, web browsing, video conferencing, light gaming \cite{19_MSFT}), while it is less critical for \hj{a} desktop system. 
Therefore, the main desktop energy-efficiency benchmarks\hj{, such as ENERGY STAR \cite{energy_star,energy2014energy} and Intel Ready Mode Technology \cite{intel_rmt,bolla2016assessing} efficiency benchmarks,} are related to reducing energy consumption once the processor is fully idle. \hj{The average power consumption needs \hja{of} such benchmarks} can be met with package $C7$ \hj{state}. 
Therefore, to reduce motherboard component cost and validation\footnote{Package $C10$, for example, requires 1) dedicated components on the motherboard to turn off IO signals, 2) a special flow to move the CPU cores' context to a dedicated area in DRAM, and 3) migrating the processor \hj{wake-up timers} to \hj{the} chipset to enable turning off the processor's crystal  clock\cite{haj2020techniques}.} effort, desktop \hj{systems} \hj{are designed} to support only up to package $C7$ \hj{state}.  
Second, \hja{supporting different features for different market segments is essential for product \hjb{specialization and cost efficiency}. For example, such differentiation prevents} laptop manufacturers from using a processor that is dedicated to \hj{the} desktop market segment\hja{, which is significantly cheaper \hj{than} a mobile processor with equivalent TDP,} to build laptop devices. 

Since the CPU core's voltage regulator is \hj{turned} on in \hj{the} package $C7$ \hj{state} (\hj{as shown} in Table \ref{tab:p-states}) and the power-gates are bypassed in \tech, the power \hj{consumption} of package $C7$ is significantly (more than $3\times$) higher \hj{in \tech} than \hj{in} the baseline due to the additional leakage power of the ungated CPU cores. To mitigate this issue, we extend the desktop systems with \hj{the} package $C8$ \hj{state} \cite{intel_skl_dev,haj2018power,haj2020flexwatts}\hj{:} a deeper (lower power but with higher \hj{entry}/exit latency) package C-state in which the voltage regulator of the CPU cores is off, as shown in Table \ref{tab:p-states}. 
\hj{This reduces the CPU cores' leakage power and saves} even more power in the uncore compared to \hja{the} package $C7$ \hj{state}. 
%\section{Implementation, \hy{Overhead, and Drawbacks}}

\section{Implementation and Hardware Cost}
% \begin{sloppypar}
{\tech}' three key components are implemented within the \om{Intel Skylake} {SoC}~\agya{\cite{11_fayneh20164,yasin2019metric,anati2016inside}}. %\agycomment{@Jawad: please check}
% \end{sloppypar}

First, {\tech}~requires the implementation of different package\om{s} for \om{the} processor segment with power-gates (i.e., Skylake-H, used for high-end mobile system) and the processor segment that bypasses the power-gates (i.e., Skylake-S, used for high-end desktop system\om{s}). Typically, these two processor segments \oma{already} have different package\om{s}: a land grid array (LGA \cite{kujala2002transition}) package for Skylake-S and a ball grid array (BGA \cite{guenin1995analysis}) package for Skylake-H. 

Second, \om{we implement} \tech's power management flows \om{i}n firmware\footnote{A fully-hardware implementation is also possible. However, such power management flows are normally error-prone and require post-silicon tuning. As such, most of the power management flows are implemented within the power-management firmware (e.g., Pcode~\cite{gough2015cpu}).} to enable the \emph{hybrid} architecture \om{o}n the processor die. \tech operates in one of two modes based on \om{a} silicon fuse~\cite{kulkarni2009high} value:
1)~\emph{bypass  mode}, which bypasses the power-gates to increase the voltage and frequency of the CPU cores, and~2)~\emph{normal mode}, which uses the power-gates to reduce the leakage power of the CPU cores. 
The additional firmware code to support this flow is approximately $0.3KB$\agy{, which is} less than $0.004\%$ of Intel Skylake's die area~\cite{21_doweck2017inside}.    

% \begin{sloppypar}
\om{Third}, \tech requires the implementation of a deeper package C-state (i.e., package $C8$) for desktop systems (i.e., Skylake-S). The package C-state power-management hardware and firmware flows are already implemented in the baseline used for mobile \om{systems}~\cite{haj2020techniques, intel_skl_dev}. Therefore, \om{we expect} no additional cost \om{for this third component}.
% \end{sloppypar}

% \agycomment{I think putting Overhead and Drawbacks to the title is not very preferable. I'd add it as a subtitle or maybe even as some bold paragraph name like \textbf{Limitations.} DarkGates achieves X and Y with three limitations: 1)~reliability, 2)~reduction in power budget, and 3)~additional design cost for a new package. We elaborate on these limitations in Sections~X, Y, and~Z, respectively.}\agycomment{I'm not sure about the word limitation as well. It's better to find something milder than overhead and drawback...} 
\begin{sloppypar}
{Like many other architectural optimizations, \mbox{\tech} \om{also} has \textbf{tradeoffs} and \textbf{drawbacks}. 
First, as explained in Sec.~\mbox{\ref{sec:pm_alg}}, \mbox{\tech} can affect the lifetime reliability of the processor. 
Second, the proposed mechanisms of \mbox{\tech} can degrade the performance of power-limited processor scenarios where few cores are active (e.g., computer graphics). \agy{This is because} the additional leakage of the inactive cores reduces the power budget allocated to the graphics engines (as shown in Fig. \mbox{\ref{fig:gfx_res}}). Third, \mbox{\tech} requires separate designs for the target segments. Although our baseline system have two separate packages for the target processor (i.e., Skylake-H and Skylake-S, discussed in Sec. \mbox{\ref{sec:pg_bypassing}}), processor vendors that do not have two packages in the baseline architecture need to build two packages to implement \mbox{\tech}.}
\end{sloppypar}

\section{Evaluation Methodology}
\label{sec:method}

% \agycomment{I think this first paragraph should come later. It can be already unnecessary actually since all of this information is provided again in two methodologies.} 
% We evaluate \tech{} using 1)~real systems that employ Intel Broadwell~\cite{5_nalamalpu2015broadwell} and Intel Skylake~\cite{Skylake_D,Skylake_H} processors and~2)~an in-house power delivery network simulator \cite{aygun2005power,ketkar2009microarchitecture}.
% \agycomment{the same comment here about citing an in-house simulator}

We use two distinct methodologies {to} 1)~collect motivational data{, demonstrating \om{\tech{}\agy{'s}} potential benefits on our target processor's (Skylake) predecessor (Broadwell)} and~2)~evaluate \tech{} on Skylake.
The reason is that we would like  to demonstrate the potential benefits of \tech on the previous generation processor (i.e., Broadwell) of our target Skylake processor, before we implement it in the  Skylake processor. 

% To collect the  motivational data we use an in-house power delivery simulator \cite{aygun2005power,ketkar2009microarchitecture}.
% We evaluate \tech using two real systems that employ Intel Skylake~\cite{21_doweck2017inside} high-end desktop (i.e., Skylake-S \cite{Skylake_D}) and high-end mobile (i.e., Skylake-H \cite{Skylake_H}) processors.

\noindent \textbf{Methodology for Collecting Motivational Data.}
We use a Broadwell-based system~\cite{5_nalamalpu2015broadwell} to collect motivational data that shows the potential performance benefits of increasing CPU \om{core} frequency when \om{we} increas\om{e} the effective voltage (i.e., by reducing the voltage guardband).

\begin{sloppypar}
To collect the motivational data of the potential impedance improvement \om{with} Power-gate \oma{B}ypassing, {we model both the baseline (i.e., with power-\om{gate}\oma{s enabled}) and \tech{} (i.e., with \agy{\oma{P}ower-gate \oma{B}ypassing}) using} an in-house power delivery network simulator (similar to~\cite{aygun2005power,shekhar2016power}).
% We model the baseline (i.e., with power-gates) and the \tech (i.e.,  with power-gates bypassing) using the simulator.   
{We create t}he model directly from the layout files of the package and the motherboard. {We use} a voltage regulator (VR) model, attached to the motherboard to allow time domain simulations. Each processor die is configured as a dynamic current load.
\end{sloppypar}
% For a frequency domain analysis, the VR is replaced by a power supply with an $RL$ element in series. The $R$ represents the static loadline of the VR and the $L$ is proportional to the bandwidth of the VR. A dynamic voltage source is attached in place of the silicon die stimulus. The model is then simulated across a wide range of frequencies by sweep over the frequency range. The resulting voltage, which is proportional to impedance, is monitored on each of the die nodes.

\noindent \textbf{Methodology for  Evaluating \tech.}
We implement \tech on the Intel Skylake \cite{21_doweck2017inside} die that targets  high-end desktop (i.e., Skylake-S \cite{Skylake_D}) and high-end mobile (i.e., Skylake-H \cite{Skylake_H}) processors. Table~\ref{tbl:sys_setup} shows the major system parameters. For our baseline and \tech measurements we use the Skylake-H (mobile)  and Skylake-S (desktop), respectively.

\begin{table}[h]
\centering
\caption{P\om{arameters of Evaluated Systems}}
\label{tbl:sys_setup}
\resizebox{\linewidth}{!}{%
\begin{tabular}{cl}
\hline
\multirow{7}{*}{\om{Processors}} & \multicolumn{1}{l}{\om{i7}-6700K \cite{i7_6700K} Skylake-\om{S}}\\
                                            & i7-6920HQ \cite{i7_6920HQ} Skylake-\om{H}                                                                                                                   \\
                                            & \begin{tabular}[c]{@{}l@{}}CPU Core Frequencies: 0.8--4.2GHz\\
                                    Graphics Engine Frequencies: 300--1150MHz\\        
                                    L3 cache (LLC): 8MB\\ Thermal Design Point (TDP): 35--91W\\ Process technology {node}: 14nm\end{tabular} \\ \hline
\multicolumn{1}{l}{\multirow{2}{*}{Memory}} & DDR4-2133 \cite{ddr4_ref}, \om{no ECC},                                                                                                                          \\
\multicolumn{1}{l}{}                        & dual-channel, 32GB capacity                                                                                                                                  \\ \hline
\end{tabular}
%  \vspace{-5pt}
}
\end{table}

% \agycomment{do all processor configurations apply for both Skylake-S and Skylake-H? Does it make sense to split them in Table II as Skylake-S Configurations and Skylake-H configurations?}

% Fig.~\ref{fig:avgp_perf_energy} shows an example for the average-power breakdown.  
\noindent \textbf{Configuring the Processor.} We use Intel's In-Target Probe (ITP) \cite{intel_itp} silicon debugger tool that connects to an Intel processor through {the} JTAG port~\cite{williams2009low}. We use ITP to configure \om{processor} control and status registers (CSRs) and model specific registers (MSRs). For example, we use the ITP to configure the TDP to multiple values between $35W$ to $91W$. 
\hj{For more detail, we refer the reader to the Intel ITP manual \cite{itp_2020,itp_xlsoft} and to our recent prior work \cite{hajsysscale}.}

% \omcomment{Can we provide a picture of the measurement setup? That is always good.
% Also, please reference our papers that used a similar setup and say so.}
% \agycomment{I think VAMPIRE's~\cite{ghose2018vampire} setup might seem similar to what we have here. However, it's much simpler in VAMPIRE and the sampling rate is significantly lower, i.e., in the order of KHz.}

\noindent \textbf{Power Measurements.} We measure power consumption when running energy-efficiency benchmarks by using a National Instruments Data Acquisition (NI-DAQ) card (NI-PCIe-6376~\cite{NIDAQ}), whose sampling rate is up to 3.5~{M}ega-samples-per-second (MS/s). Differential cables transfer multiple signals from the power supply lines on the motherboard to the {NI-}DAQ card in the host computer that collects the power measurements. By using NI-DAQ, we measure {power on} up to 8 channels {simultaneously}. We connect {each} measurement channel to one  voltage regulator of the processor. The power measurement  accuracy of the NI-PCIe-6376 is $99.94\%$. \oma{Our prior works~\cite{hajsysscale,haj2021ichannels} provide more detail on this experimental setup.}

\noindent \textbf{Workloads.} We evaluate {\tech} with three classes of workloads that are widely used for evaluating client processors. 
First, to evaluate  CPU core performance, we use  the SPEC CPU2006 benchmarks~\cite{SPEC2018} {and} use the SPEC CPU2006 benchmark score as the performance metric. 
Second, to evaluate  computer graphics performance, we use the 3DMARK benchmarks~\cite{17_3DMARK} {and} use frames per second (FPS) as {the} performance metric. 
Third, to evaluate the effect {of} \tech on energy efficiency, we {measure the average power consumption of} two workloads that are typically used to evaluate energy consumption of desktop \om{systems} (\om{e.g.}, the Skylake-S){:}  
% We use average power consumption as energy efficiency evaluation metric.
1)~\emph{ENERGY STAR} is a program that promotes energy efficiency~\cite{energy_star,energy2014energy}.
An important criterion {of ENERGY STAR} is that a \om{system} must automatically enter into a low power mode, defined as off, sleep, long\_idle, short\_idle, {when it is} idle. The \om{depth} of the low power mode is determined based on the \agy{idle} period of the \agy{system}. 
The energy consumption limit values are calculated using a formula that is based on the residency \om{in} each power state and the power \om{consumption of} each state.  
% \omcomment{Intel RMT does not sound like a workload description. There is a mismatch here. What is the workload?}
%\agycomment{Maybe we should name this part as Tests or Benchmarks instead of Workloads. The structure would be 1)~CPU Performance Tests (SPEC), 2)~Graphics Performance Tests (3DMARK), and 3)~Light Workload/Idle Tests. In 3), we use ENERGY STAR to measure the energy consumption in X,Y,Z states; and Intel RMT to measure the energy consumption in another set of X,Y,Z states. Does that make sense?}
\hjb{2) An idle platform workload that places the platform into \emph{Ready Mode}  enabled by Intel's ~\emph{Ready Mode Technology} (RMT \cite{intel_rmt,bolla2016assessing,intel_rmt2}).}\footnote{\hjb{Intel Ready Mode Technology (RMT) provides an alternative to the traditional desktop sleep state, such as suspend states S3 (suspend to RAM) and S4 (suspend to disk)~\cite{haj2018power,gough2015energy,tu2015atom}. A similar feature exists in mobile devices, called Connected-Standby \cite{haj2020techniques,haj2012transferring,haj2013connected}.}} 
A modern desktop \om{system} enters into \hjb{Ready Mode} during idle periods, in which it operate{s} at \om{a} low power state (e.g., package $C7$ \cite{rmt_anandtech}) to reduce energy consumption while remaining connected to a communication network for usability (e.g., for email notifications and phone calls). Typically, ${\sim}99\%$ of the time, the platform is idle (e.g., in package $C7$ \om{state}) and consumes few hundreds of milliwatts \cite{haj-micro-2021,deval2015power,intel_rmt2}. In the remaining ${\sim}1\%$ of the time, the platform is active (in package $C0$ \om{state}) and consumes a few watts \cite{haj2020techniques,intel_rmt,bolla2016assessing}. 
%\omcomment{Intel RMT still does not wound like a benchmark based on this description. Yet, we called it a benchmark. What is wrong?}

%\agycomment{I'm confused. We were talking about workloads. We say that we use two workloads to evaluate the effect of energy efficiency on something not defined. Then, we talked about two things that look like evaluation tools rather than workloads...}
\section{Evaluation}\label{sec:eval}
%Our baseline system is running with fixed frequencies for high-bandwidth interconnect and memory subsystem with fixed static power-budget (i.e. not dynamically shared with cores or graphics). Whereas, when running with \tech, the frequencies of the high-bandwidth interconnect and memory subsystem is dynamically adjusted based on the demand from the multiple-domains sharing these resources. When high-bandwidth interconnect and memory subsystem run at lower frequencies (and voltage) these domains consumes less average power, thereby the processor runs below TDP. As such, the spared average power is used by the power management unit (PMU) to increase the frequency of the other domains, for example increase the cores frequency by $200MHz$ or increase graphics frequency by $100MHz$. The performance impact on the running workload, when raising the core or graphics frequency depends on the scalability\footnote{Performance-scalability is performance and frequency correlation; a ratio of 1 means doubling frequency results in doubling performance} of the running workload.    

 \begin{figure*}[!ht]
  \begin{center}
  \includegraphics[trim=0.7cm 0.7cm 0.7cm 0.7cm, clip=true,width=0.99\linewidth,keepaspectratio]{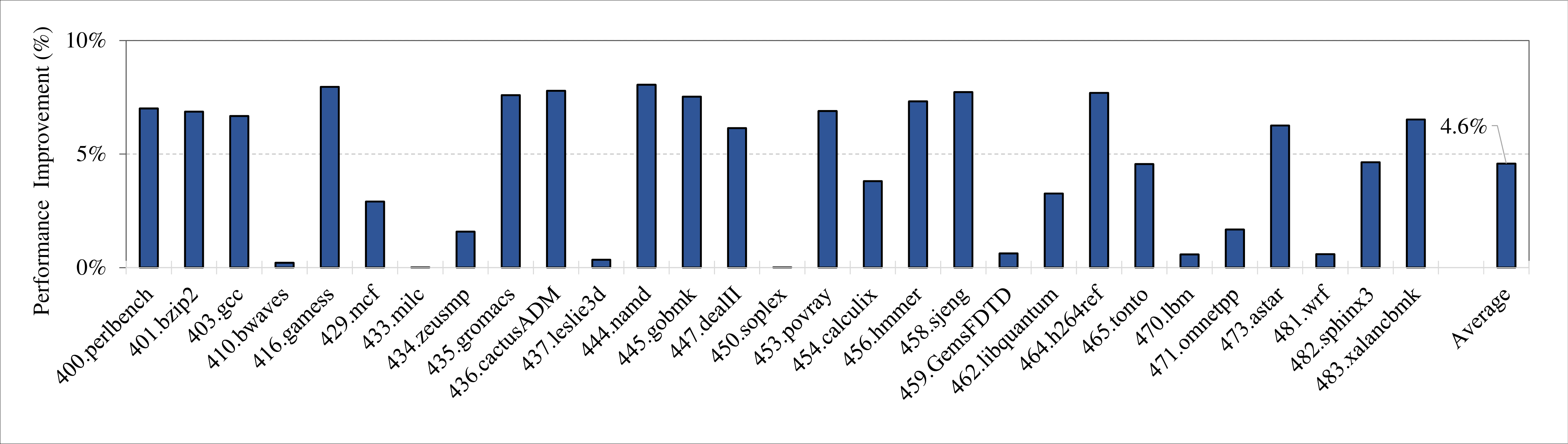}\\
%  \vspace{-15pt}
  \caption{Performance improvement of {\tech} when it is implemented in \om{the} Intel Skylake-S processor on SPEC CPU2006 workloads compared to \om{the} baseline \om{Intel} Skylake-H processor with enabled power-gates\om{;} both processors \om{have a} TDP of $91W$.}\label{fig:core_res}
  \end{center}
%  \vspace{-20pt}
 \end{figure*}

We present performance and average power benefits %\footnote{Energy efficiency, {i.e.,} energy-delay-product (EDP) {\cite{gonzalez1996energy}}, improves  proportionally to {performance or} average power {consumption}, since \tech improves either performance within a fixed power budget or average power consumption within a fixed performance requirement.}
obtained with {\tech} when it is implemented in \om{the} Intel Skylake-S~\cite{Skylake_D} processor compared to Skylake-H~\cite{Skylake_H} \om{baseline} that ha\om{s} power-gates enabled. 
We evaluate three workload categories: CPU (Sec. \ref{res:cpu}), graphics (Sec. \ref{res:graphics}), and energy efficiency workloads (Sec. \ref{res:battery}). 
% \agycomment{Are these three workloads the workloads that we explain in the previous section? If so, we need to work on the consistency across sections if we change terminology.}
%We also analyze  sensitivity to different SoC thermal-design-power (TDP) levels and DRAM frequencies (Sec. \ref{res:sens}).

\subsection{Evaluation of CPU Workloads}
\label{res:cpu}
 %% FigRemoved

%We compare {\tech} to MemScale~\cite{deng2011memscale} and CoScale~\cite{deng2012coscale}, the most relevant works to our technique. We compare the average power reduction of each one of the three techniques when running SPEC CPU2006 workloads. We also evaluate the performance impact of {\tech} with SPEC CPU2006 workloads.
Fig. \ref{fig:core_res} reports the performance improvement of {\tech} when it is implemented in \om{the} Intel Skylake-S processor on SPEC CPU2006 base \hj{(single core}) workloads over the baseline Skylake-H processor with enabled power-gates when both processors operate at the\om{ir} highest attainable CPU core frequencies within a TDP of $91W$.
We make two key observations. %\todo{Double check the number of observations.}

First, {\tech} improves real system performance by up to $8.1\%$ ($4.6\%$ on average). This result is significant as it is obtained on a real \om{Intel Skylake-S} system. 

%The average power reduction is because MemScale scales only memory, while CoScale scales  memory and CPU cores. %For fair comparission, we assume that  CoScale and MemScale can utilize the saved average power budget to redistribute power budget across domains (although this is not implements in both techniques).

Second, the performance benefit of {\tech} \om{is positively} {correlate\om{d} with} the \emph{performance scalability}\footnote{We define performance scalability of a workload with respect to CPU frequency as the performance improvement the workload experiences with unit increase in frequency, as described in \cite{yasin2017performance,haj2015doee}.}
%\footnote{{We define performance scalability of a workload with respect to CPU frequency as the improvement the workload gets in performance with an increase in frequency.}}
of the running workload with CPU frequency. Highly-scalable workloads (i.e., those bottlenecked by CPU core \om{frequency}, {such as \emph{416.gamess} and \emph{444.namd}}) \om{experience} the highest performance gains. In contrast, workloads that are heavily bottlenecked by main memory, such as \emph{410.bwaves} and \emph{433.milc}, have almost no performance gain.

We conclude that \tech significantly improves CPU core performance by reducing the voltag\om{e} guardband \om{with Power-gate \oma{B}ypassing}, which improves the V/F curve \om{and leads to} higher CPU core frequency.

\begin{figure}[!ht]
  \begin{center}
  \includegraphics[trim=0.7cm 0.7cm 0.7cm 0.7cm, clip=true,width=0.85\linewidth,keepaspectratio]{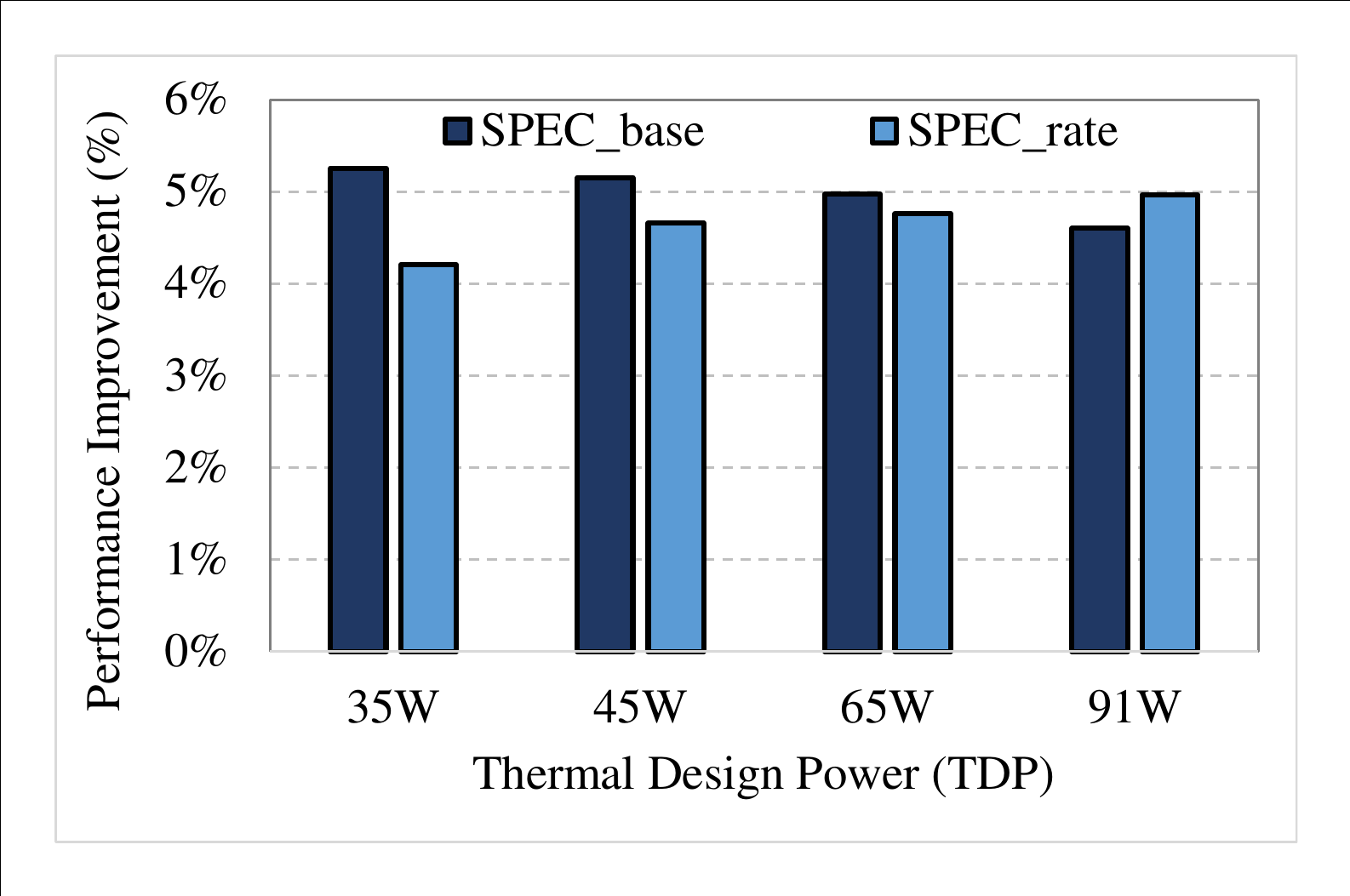}\\
  \vspace{-0.5em}
  \caption{Average \om{p}erformance improvement of {\tech} when it is implemented in \om{the} Intel Skylake-S processor on SPEC CPU2006 (base and rate) workloads compared to \om{the} baseline \om{Intel} Skylake-H processor with enabled power-gates for multiple TDP levels.}\label{fig:core_res_tdps}
  \vspace{-0.5em}
  \end{center}
%  \vspace{-15pt}
 \end{figure}

Fig. \ref{fig:core_res_tdps} reports the average performance improvement of {\tech} when it is implemented in \om{the} Intel Skylake-S processor on SPEC CPU2006 base \hj{(single core)} and rate \hj{(all cores)} workloads over the baseline Skylake-H processor with enabled power-gates when both processors operate at the highest attainable CPU core frequencies \om{at} multiple TDP levels ($35W$, $45W$, $65W$, and $91W$). We make three key observations.

First, {\tech} improves \om{the average} real system performance of SPEC base/rate \om{benchmarks} by  $5.3\%$/$4.2\%$, $5.2\%$/$4.7\%$, $5.0\%$/$4.8\%$, and $4.6\%$/$5.0\%$ for $35W$, $45W$, $65W$, and $91W$ \om{TDP}, respectively. This result is \om{also} significant as it is obtained on a real \om{Intel Skylake-S} system. 

% \begin{sloppypar}
Second, the average performance improvemen\om{t} of SPEC\_base benchmarks decreases as the T\om{D}P level increase\om{s}. The reason is that \om{at} a low TDP (e.g., $35W$)\om{,} the processor is more thermally constrained and run\om{s} at \om{a} lower frequency than \om{at a} higher TDP (e.g., $91W$). \om{T}herefore, the relative increase \om{in} frequency, \om{at} steps of $100MHz$ granularity until reaching the TDP limit, is higher \om{at a} lower TDP.
% \end{sloppypar}

% \begin{sloppypar}
Third, the average performance improvements of SPEC\_rate benchmarks increases as the T\om{D}P level increase\om{s}. The reason is that a high TDP is $V_{max}$-constrained while \om{a} low TDP \om{is} thermally constrained\om{. T}herefore, \agy{a} low TDP processor exceed\om{s} the TDP limit faster once all cores \om{operate} at \om{an} increased frequency compared \om{to} a high TDP processor (e.g.,  $91W$) that can increase the frequency of all cores to the maximum attainable frequency  with the improved $V_{max}$ (due to the reduc\om{ed} voltage guardband) without exceeding the thermal limit. 
% \end{sloppypar}

% \begin{sloppypar}
We conclude that \tech significantly improves CPU core performance by reducing the voltage guardband \om{with Power-gate} bypassing, which improves the V/F curves \om{and leads to} higher CPU core frequency for both thermally\om{-}constrained and $V_{max}$-constrained systems.
% \end{sloppypar}

% \vspace{5pt}

\subsection{Evaluation of Graphics Workloads}
\label{res:graphics}

\begin{sloppypar}
Typically, the performance of a graphics  workload  is highly scalable with the graphics engine frequency. When running graphics workloads, the power budget management algorithm (PBM \cite{lempel20112nd,rotem2015intel}) of the PMU normally allocates only $10\%$ to $20\%$ of the compute domain power budget to the CPU cores, while the graphics engine consumes the rest of the power budget~\cite{rotem2011power,rotem2012power,rotem2013power}. 
For a client system, while running a graphics workload, one of the the CPU core\om{s} normally runs (e.g., runs the graphics driver) at the most energy\om{-}efficient frequency $Pn$~\cite{haj2018power} (i.e., the maximum possible frequency at the minimum functional voltage ($V_{min}$)) while the other cores are idle and power-gated. Since the power-gates are bypassed in a system with \tech, the additional leakage of the inactive cores (\om{i.e.,} three CPU cores in a four\om{-core} processor) reduces the effective power budget allocated to graphics engines, which can reduce the \om{graphics} performance of \om{a} thermally\om{-}constrained system.
\end{sloppypar}

%\vspace{-5pt}

 \begin{figure}[!ht]
  \begin{center}
  \includegraphics[trim=.99cm .8cm .8cm .8cm, clip=true,width=\linewidth,,keepaspectratio]{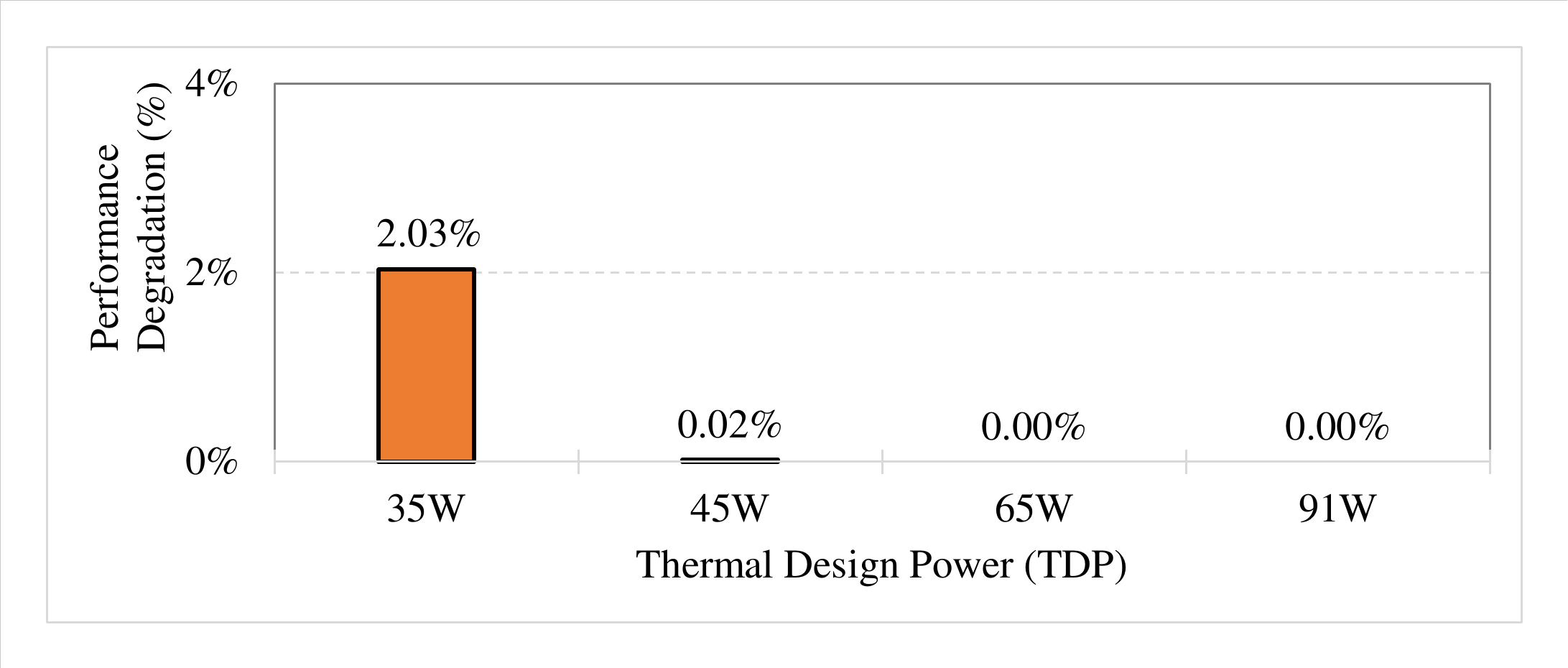}\\
%  \vspace{-10pt}
  \caption{\om{A}verage performance degradation \om{of \tech{}} \oma{over} \om{the} baseline system across different TDP \om{level}s when running \om{3DMark} graphics workloads.}
  \label{fig:gfx_res}
  \end{center}
%  \vspace{-5pt}
 \end{figure}
 
Fig.~\ref{fig:gfx_res} shows the average performance degradation of {\tech} compared to \om{the} baseline system across different TDP \om{level}s  
%and the performance improvement of {\tech} \sr{no performance improvements for memscale and coscale?} 
when running \om{the} 3DMark~\cite{17_3DMARK} graphics workloads. 
We make two key observations.

% \begin{sloppypar}
First, \tech provides the same system performance \om{for} 3DMark for TDP \om{levels equal to or} higher than $45W$. \om{Graphics} performance is not affected with the additional power \om{spent on the leakage of idle} cores since graphics workloads in these systems are not limited by thermal \om{constraints}. 
% In fact, \om{Intel} Skylake-S with \om{a} TDP \om{level from} $45W$ to $91W$ operate\om{s} at $350MHz$ base frequency, which mean\om{s} that these workloads are not thermally limited.
% \omcomment{This sentence is odd. We are talking about workloads but giving frequency for TDP levels. Why is frequency constant. What is this frequency for? The graphics engine? Poor writing. Fix this sentence for sure. }
% \end{sloppypar}

 %{\tech} obtain the power budget savings by the holistic DVFS it apply to SoC domains based on their dynamic demand. 

%% FigRemoved

Second, for \om{a} TDP \om{level of $35W$}, \tech \om{leads to} only $2\%$ performance degradation \om{in graphics workloads}. 
%The performance improvement obtained by 
{\tech} reduces the  graphics performance for a  system with $35W$ TDP because this system is thermally limited\om{. H}ence\om{,} the additional leakage power of the \om{idle CPU} cores forces the PBM to reduce the frequency of the graphics engine to keep the system within the TDP limit.

% Third, \sr{not sure if this is an observation. It seems like an explanation or note. I think this is more general description and not restricted to SysScale!. We can move all of this to the motivation or some where else? I think we need to explain why the SysScale performance in Fig. 8 is different for different workloads? based on what?} 
% In mobile system, the graphics engine frequency is normally thermally constrained to keep the processor power within TDP. For example the processors described in Table \ref{tbl:sys_setup} are capable of running opportunistically (turbo \cite{rotem2013power}) above $800MHz$, but their sustainable frequency is $300MHz$. 

%\sr{Please fix the conclusion based on the results in this subsection}
We conclude that the reduced graphics engine power budget due to the additional \om{leakage power of idle} CPU cores can slightly degrade the performance of graphics workloads in thermally\om{-}limited systems\om{, but it is not a main concern in many real systems that are not thermally-limited.}
% \sr{This example is misplaced. Why we are talking about Fig. 6 here?}For example, when raising the graphics engine frequency of the Skylake processor from $300MHz$ to $350MHz$ (without changing memory subsystem frequencies) the performance of 3DMark-Vantage \cite{17_3DMARK} improves by around $16\%$ \sr{which figure?}. While when reducing the frequency of the memory subsystem from $1.6GHz$ to $1.06GHz$, the average performance degradation of the traces is about $6\%$ as shown in Fig. \ref{fig:predict_perf_impact}(h).      

\subsection{Evaluation of  Energy Efficiency Workloads}
\label{res:battery}

Unlike  CPU  and graphics workloads that always benefit from higher performance, energy efficiency workloads, such as \emph{ENERGY STAR} \cite{energy_star,energy2014energy}) and \emph{Intel Ready Mode Technology} (RMT \cite{intel_rmt,bolla2016assessing}), 
have long idle phases where the system enters into idle power states (i.e., C-states~\cite{haj2018power,acpi,gough2015cpu,haj2020techniques}). 
For example, in the RMT workload (discuss\om{ed} in Sec. \ref{sec:method}) of the baseline system (i.e., with power-gates) the package $C0$ (i.e., active)  power state residency is \om{only} ${\sim}1\%$ of the total \om{time} and the package $C7$ (i.e., idle) power state residency is ${\sim}99\%$ of the total time. Since \tech bypasses the power-gates, package $C7$ power \om{would} significantly increase due \om{to} leakage power \om{consumed by the idle cores}. Therefore\om{,} \tech uses package $C8$ instead of $C7$ to keep the average power of these workloads within the target limits.    

\begin{figure}[!h]
  \begin{center}
  \includegraphics[trim=.8cm .8cm .8cm 0.8cm, clip=true,width=1.0\linewidth,keepaspectratio]{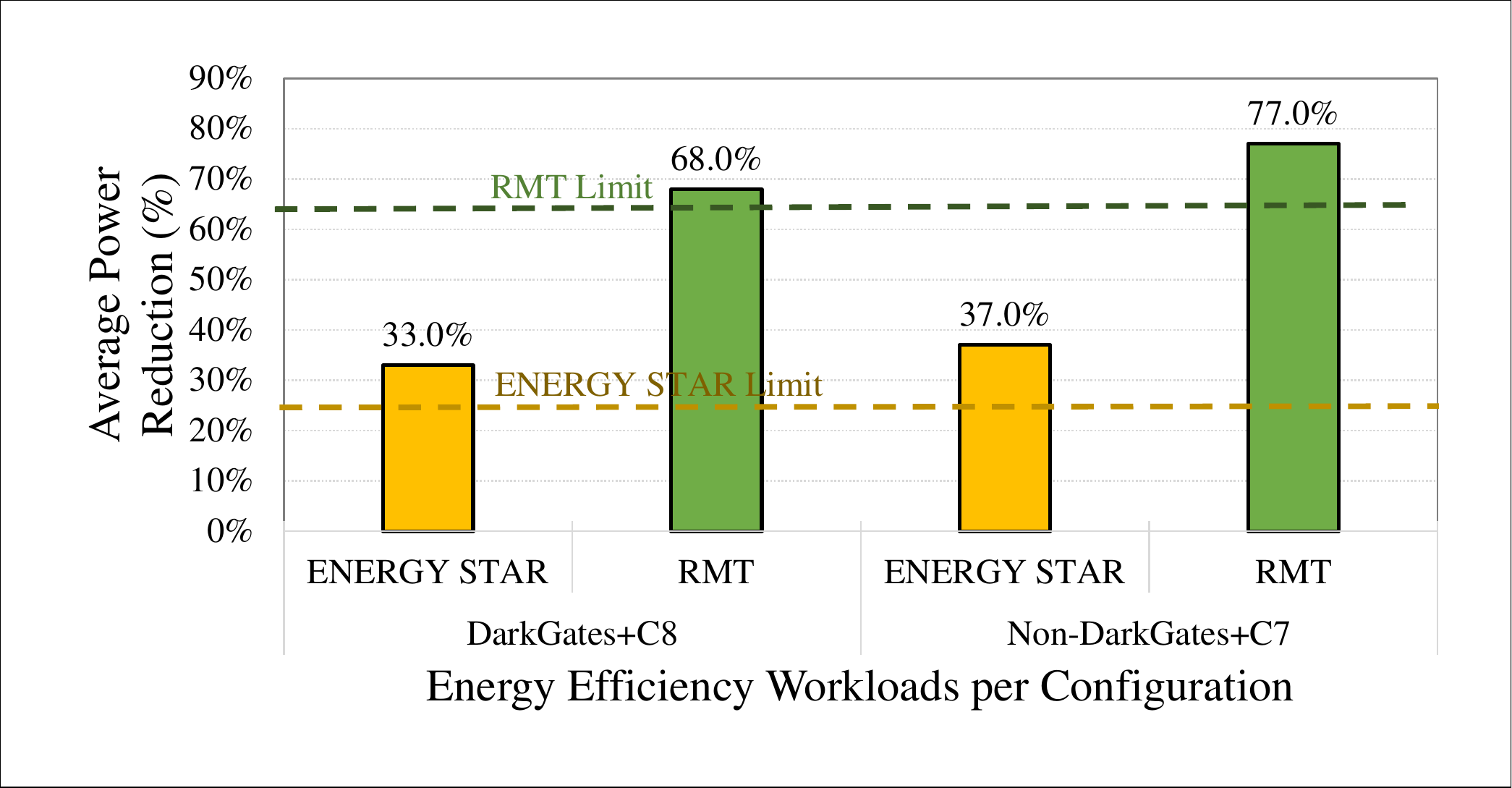}
  \caption{\om{A}verage \om{processor} power reduction when running two energy efficiency workloads, ENERGY STAR and RMT on \hy{1)}~the \tech system \hy{\om{at package}~$C8$ \om{power state} (\mbox{\tech}$+C8$), and 2)~a system without \mbox{\tech} \om{at package}~$C7$ \om{power state} (Non-\mbox{\tech}$+C7$) compared to} the baseline system \hy{\om{that} includes \mbox{\tech} \om{at} $C7$ (\mbox{\tech}$+C7$)}.}\label{fig:avg_power_res}
  \end{center}
%  \vspace{-5pt}
 \end{figure}

% \agycomment{Alternative: Fig.~\ref{fig:avg_power_res} shows the average \om{processor} power reduction \agy{on \om{Intel Skylake-S}} \agy{for} the two energy-efficiency workloads (i.e., ENERGY STAR and \agy{Intel} RMT) for two systems: 1)~\tech system (where power-gates is bypassed, and the deepest package C-state is $C8$ \hy{(denoted by \mbox{\tech}$+C8$}) and 2)~a \om{baseline} system without \mbox{\tech}\om{, and} the deepest package C-state is $C7$ (denoted by Non-\mbox{\tech}$+C7$). }
Fig.~\ref{fig:avg_power_res} shows the \om{Intel Skylake-S} average \om{processor} power reduction when running the two energy-efficiency workloads, ENERGY STAR and RMT, on \hj{two systems:} 
\hy{1)} the \tech system, where power-gates \hj{are} bypassed, and the deepest package C-state is $C8$ \hj{(denoted by \mbox{\tech}$+C8$}), \hy{and 
2) a system without \mbox{\tech}\om{, and} the deepest package C-state is $C7$ (denoted by Non-\mbox{\tech}$+C7$),} \hy{when compared to} the  \hj{baseline} system that includes \oma{\tech but limits} the deepest package C-state to $C7$ \hy{(denoted by \mbox{\tech}$+C7$)}.

We make \hy{three} key observations.
First, our \om{proposed} {\tech} \om{system} \hy{(i.e., \mbox{\tech}$+C8$)} reduces the average power consumption of ENERGY STAR and RMT by  $33\%$ and $68\%$, respectively, on \om{the} real \om{Intel Skylake-S} system, compared to the baseline \om{where} we limit the deepest package C-state to $C7$ \hy{(i.e., \mbox{\tech}$+C7$)}.

Second, the baseline system (i.e., {\tech}$+C7$) does not meet the target average power limit for both workloads. Applying package $C8$ is essential to meet the target average power \om{limit} since $C8$ turns off the CPU core\om{'}s voltage regulator, \om{which greatly} reduc\om{es} the leakage power of the ungated CPU cores. 

% \begin{sloppypar}
\hy{Third, the system without \mbox{\tech} \om{at} the deepest package C-state of $C7$ (i.e., Non-\mbox{\tech}$+C7$) shows lower average power consumption compared to the system with \mbox{\tech} \om{at} the deepest package C-state of $C8$ (i.e., \mbox{\tech}$+C8$). The higher average power of \mbox{\tech}$+C8$ compared to Non-\mbox{\tech}$+C7$ is mainly because \mbox{\tech}$+C8$ has higher power consumption \om{at} the power states in which some of the cores are idle, such as $C0$. These cores consume leakage power in \mbox{\tech}$+C8$\om{, but} they are power-gated in Non-\mbox{\tech}$+C7$}.  
We conclude that for energy efficiency  workloads, which have fixed performance requirements, applying {\tech} with package $C8$ \om{state} significantly reduces the average processor power consumption\om{, thereby} meeting the target average power requirements of \om{the} energy efficiency standards.

\section{Related Work}
To our knowledge, {\tech} is the first \jk{hybrid power-gating architecture for} different processor \jk{market} segments \jk{that increases} the performance of systems constrained by the maximum \jk{attainable} CPU core frequency (i.e., \emph{$F_{max}$-constrained}), such as high-end  desktop systems. \jk{There} are many prior works that discuss the  reduction \cite{cho2016postsilicon,kim2008silicon,cho2015aging,leng2015safe,bacha2013dynamic,von1990voltage,lefurgy2013active,bertran2014voltage,papadimitriou2017harnessing,thomas2016core,miller2012vrsync,leng2020predictive,lefurgy2011active,ernst2003razor,brooks2001dynamic} and characterization \cite{joseph2003control,powell2003pipeline,gupta2007understanding,miller2012vrsync}  of various design voltage \jk{guardbands}.  Other works focus on system impedance characterization and optimization \cite{engin2010efficient,shekhar2016power}. \jk{All these} works are either orthogonal to \tech (i.e., can be applied with \tech) or they do not propose a practical and hybrid mitigation \jk{approach to reduce} power-gates' impedance. 

\noindent {\textbf{Voltage Guardband Reduction.}}
Many prior works propose multiple techniques to reduce voltage guardband \cite{cho2016postsilicon,kim2008silicon,cho2015aging,leng2015safe,bacha2013dynamic,von1990voltage,lefurgy2013active,bertran2014voltage,papadimitriou2017harnessing,thomas2016core,miller2012vrsync,leng2020predictive,lefurgy2011active,ernst2003razor,das2008razorii,das2006self}.
\jk{These works} can be categorized into two types. The first category reduces the voltage guardband while the processors continue to function correctly \cite{leng2020predictive,lefurgy2011active,haj2015compiler}, whereas the second category tolerates timing speculation errors with the aid of an error detection and recovery mechanism \cite{ernst2003razor,das2008razorii,das2006self}. 
%Similar to the former scenario, \tech assumes that the voltage drop guardband reduction does not introduce timing error.
These mechanisms optimize voltage guardband using hardware and/or software sensors to reduce the operating margin for energy \jk{savings}. \jk{Multiple} of these guardband reduction mechanisms are already applied in the Skylake processor~\jk{\cite{Skylake_H, Skylake_D,anati2016inside,tam2018skylake,yasin2019metric}}. \jk{\tech} can be applied orthogonally to these mechanisms since it physically optimizes the system impedance by bypassing the power-gates and sharing the power delivery resources on the package.

\begin{sloppypar}
\noindent {\textbf{Voltage Guardband Characterization.}} 
Several prior works \jk{use simulation} to study voltage noise in single-core \cite{joseph2003control,powell2003pipeline,brooks2001dynamic} and \jk{multi-core} \cite{gupta2007understanding,miller2012vrsync} CPUs. Other prior works conduct measurement-based \jk{studies} of voltage noise in CPUs \cite{bertran2014voltage,kim2012audit,reddi2009voltage,reddi2010voltage,lefurgy2011active}. In our work, we use an in-house power delivery network simulator \cite{aygun2005power,ketkar2009microarchitecture} to characterize the system impedance and the voltage guardband. Our model is created directly from the layout files of the package and the motherboard as well as measured data from the previous \jk{Intel} processor generation \hj{(Intel Broadwell \cite{5_nalamalpu2015broadwell})}.
\end{sloppypar}

\begin{sloppypar}
\noindent {\textbf{System Impedance Characterization and Optimization.}}
Shekher \textit{et al.} \cite{shekhar2016power} discuss different microprocessor power gating architectures and their  impact on system impedance. Engin \textit{et al.} \cite{engin2010efficient} present efficient algorithms for sensitivity calculations of power delivery network impedance to minimize the maximum deviation from the target impedance. Compared to \tech, these works do not propose practical mitigation techniques, such as \hjb{Power-gate Bypassing}.
\end{sloppypar}

\section{Conclusion}

\begin{sloppypar}
We propose \tech, the first \jk{hybrid power-gating architecture for} different processor \jk{market} segments \jk{that increases} the performance of systems constrained by the maximum \jk{attainable} CPU core frequency (i.e., \emph{$F_{max}$-constrained}), such as high-end  desktop systems.
\tech enables optimizing and customizing 
% introduces the ability to optimize and customize 
the processor package, firmware, and features based on the target processor \jk{market segment} needs by overcoming the limitations introduced by power gates. 
\tech is implemented in \jk{the Intel Skylake processor} for high-end desktops and mobile processors (i.e., Skylake-S and Skylake-H).
On a real 4-core Skylake system with integrated graphics, \tech{} improves the performance of SPEC CPU2006 workloads by up to $8.1\%$ ($4.6\%$ on average) for the \jk{highest} thermal design power (TDP) desktop system (i.e., $91W$). \tech maintains the performance 3DMark \jk{graphics} workloads for a desktop system with TDP higher than $45W$ while \jk{a} $35W$-TDP (the lowest TDP) desktop \jk{experiences} only $2\%$ performance degradation. %\agycomment{2\% what? slowdown? speedup? energy overhead? }    
{\tech} fulfills the ENERGY STAR (energy efficiency standard) and the Intel Ready Mode energy efficiency benchmark requirements.
We conclude that {\tech} is an effective approach to \jk{improving energy consumption and performance demands across high-end heterogeneous client processors.}
% in a sophisticated heterogeneous client processors
\end{sloppypar}

% use section* for acknowledgement
\section*{Acknowledgment}
\begin{sloppypar}
We thank the anonymous reviewers of HPCA 2022 for feedback. We thank the
SAFARI Research Group members for valuable feedback and the stimulating intellectual environment they provide. We acknowledge the generous gifts provided by our industrial partners: Google, Huawei, Intel, Microsoft, and VMware.
\end{sloppypar}

\balance
% trigger a \newpage just before the given reference
% number - used to balance the columns on the last page
% adjust value as needed - may need to be readjusted if
% the document is modified later
%\IEEEtriggeratref{8}
% The "triggered" command can be changed if desired:
%\IEEEtriggercmd{\enlargethispage{-5in}}

% references section

% can use a bibliography generated by BibTeX as a .bbl file
% BibTeX documentation can be easily obtained at:
% http://www.ctan.org/tex-archive/biblio/bibtex/contrib/doc/
% The IEEEtran BibTeX style support page is at:
% http://www.michaelshell.org/tex/ieeetran/bibtex/
%\bibliographystyle{IEEEtran}
% argument is your BibTeX string definitions and bibliography database(s)
%\bibliography{IEEEabrv,../bib/paper}
%
% <OR> manually copy in the resultant .bbl file
% set second argument of \begin to the number of references
% (used to reserve space for the reference number labels box)
%%%%%%%%% -- BIB STYLE AND FILE -- %%%%%%%%
\bibliographystyle{IEEEtran}
\bibliography{refs}

% Generated by IEEEtran.bst, version: 1.14 (2015/08/26)
\begin{thebibliography}{100}
\providecommand{\url}[1]{#1}
\csname url@samestyle\endcsname
\providecommand{\newblock}{\relax}
\providecommand{\bibinfo}[2]{#2}
\providecommand{\BIBentrySTDinterwordspacing}{\spaceskip=0pt\relax}
\providecommand{\BIBentryALTinterwordstretchfactor}{4}
\providecommand{\BIBentryALTinterwordspacing}{\spaceskip=\fontdimen2\font plus
\BIBentryALTinterwordstretchfactor\fontdimen3\font minus
  \fontdimen4\font\relax}
\providecommand{\BIBforeignlanguage}[2]{{%
\expandafter\ifx\csname l@#1\endcsname\relax
\typeout{** WARNING: IEEEtran.bst: No hyphenation pattern has been}%
\typeout{** loaded for the language `#1'. Using the pattern for}%
\typeout{** the default language instead.}%
\else
\language=\csname l@#1\endcsname
\fi
#2}}
\providecommand{\BIBdecl}{\relax}
\BIBdecl

\bibitem{dennard1974design}
R.~H. Dennard \emph{et~al.}, ``{Design of Ion-implanted MOSFET's with Very
  Small Physical Dimensions},'' \emph{JSSC}, 1974.

\bibitem{merritt2009arm}
R.~Merritt, ``{ARM CTO: Power Surge Could Create ``dark silicon''},'' \emph{EE
  Times}, vol.~22, 2009.

\bibitem{esmaeilzadeh2011dark}
H.~Esmaeilzadeh \emph{et~al.}, ``{Dark Silicon and the End of Multicore
  Scaling},'' in \emph{ISCA}, 2011.

\bibitem{zelikson2011embedded}
M.~Zelikson and A.~Waizman, ``{Embedded Power Gating},'' 2011, {US Patent
  7,880,284}.

\bibitem{shockley1952unipolar}
W.~Shockley, ``{A Unipolar Field-Effect Transistor},'' \emph{IRE}, 1952.

\bibitem{petrica2013flicker}
P.~Petrica \emph{et~al.}, ``{Flicker: A Dynamically Adaptive Architecture for
  Power Limited Multicore Systems},'' in \emph{ISCA}, 2013.

\bibitem{ditomaso2017machine}
D.~DiTomaso \emph{et~al.}, ``{Machine Learning Enabled Power-aware
  Network-on-chip Design},'' in \emph{DATE}, 2017.

\bibitem{rahman2006determination}
A.~Rahman \emph{et~al.}, ``{Determination of Power Gating Granularity for FPGA
  Fabric},'' in \emph{CICC}, 2006.

\bibitem{flynn2007low}
D.~Flynn \emph{et~al.}, \emph{{Low Power Methodology Manual: for System-on-chip
  Design}}.\hskip 1em plus 0.5em minus 0.4em\relax Springer Science \& Business
  Media, 2007.

\bibitem{hu2004microarchitectural}
Z.~Hu \emph{et~al.}, ``{Microarchitectural Techniques for Power Gating of
  Execution Units},'' in \emph{ISLPED}, 2004.

\bibitem{heo2002dynamic}
S.~Heo \emph{et~al.}, ``{Dynamic Fine-Grain Leakage Reduction using
  Leakage-Biased Bitlines},'' in \emph{ISCA}, 2002.

\bibitem{cho2016postsilicon}
M.~Cho \emph{et~al.}, ``{Postsilicon Voltage Guard-band Reduction in a 22 nm
  Graphics Execution Core using Adaptive Voltage Scaling and Dynamic Power
  Gating},'' \emph{JSSC}, 2016.

\bibitem{de2015fine}
V.~De, ``{Fine-grain Power Management in Manycore Processor and System-on-Chip
  (SoC) designs},'' in \emph{ICCAD}, 2015.

\bibitem{Skylake_D}
{Intel}, ``{{6th Generation Intel Processor Families for S-Platforms}},''
  online accessed Jul 2021 \url{https://intel.ly/2XVdORo}.

\bibitem{bowman2009impact}
K.~A. Bowman \emph{et~al.}, ``{Impact of Die-to-Die and Within-Die Parameter
  Variations on the Clock Frequency and Throughput of Multi-core Processors},''
  \emph{IEEE Transactions on Very Large Scale Integration (VLSI) Systems},
  2009.

\bibitem{lee2009optimizing}
J.~Lee and N.~S. Kim, ``{Optimizing Throughput of Power- and
  Thermal-constrained Multicore Processors using DVFS and Per-core
  Power-gating},'' in \emph{DAC}, 2009.

\bibitem{2_burton2014fivr}
E.~A. Burton \emph{et~al.}, ``{FIVR - Fully Integrated Voltage Regulators on
  4th Generation Intel{\textregistered} Core™ SoCs},'' in \emph{APEC}, 2014.

\bibitem{15_jakushokas2010power}
R.~Jakushokas \emph{et~al.}, \emph{{Power Distribution Networks with On-Chip
  Decoupling Capacitors}}.\hskip 1em plus 0.5em minus 0.4em\relax Springer
  Science \& Business Media, 2010.

\bibitem{intel_tdc}
{Intel}, ``{6th Generation Intel® Core™ Processor, External Design
  Specification (EDS) Addendum},'' \url{https://intel.ly/2Vthg79"}, May 2016.

\bibitem{gough2015cpu}
C.~Gough \emph{et~al.}, ``{CPU Power Management},'' in \emph{{Energy Efficient
  Servers: Blueprints for Data Center Optimization}}, 2015, pp. 21--70.

\bibitem{intel_skl_dev}
{Intel}, ``{6th Generation Intel® Processor for U/Y-Platforms Datasheet},''
  online, accessed June 2020, https://intel.ly/37rtnU7.

\bibitem{haj2018power}
J.~Haj-Yahya \emph{et~al.}, ``{Power Management of Modern Processors},'' in
  \emph{{Energy Efficient High Performance Processors}}, 2018, pp. 1--55.

\bibitem{anati2016inside}
I.~Anati \emph{et~al.}, ``{{Inside 6th Gen Intel{\textregistered} Core™: New
  Microarchitecture Code Named Skylake}},'' in \emph{HotChips}, 2016.

\bibitem{kabylake_2020}
{Wikipedia}, ``{7th Generation Intel Core Processor Family (Kaby Lake)},''
  online, accessed August 2020 \url{https://en.wikipedia.org/wiki/Kaby_Lake}.

\bibitem{coffeelake_2020}
{Wikipedia}, ``{8th and 9th Generation Intel Core Processor Family (Coffee
  Lake)},'' online, accessed August 2020
  \url{https://en.wikipedia.org/wiki/Coffee_Lake}.

\bibitem{i38121u_cannonlake}
{Intel}, ``{{Intel® Core i3-8121U Processor}},'' online, accessed Aug 2020
  \url{https://ark.intel.com/content/www/us/en/ark/products/136863/intel-core-i3-8121u-processor-4m-cache-up-to-3-20-ghz.html}.

\bibitem{anandtech_ipc}
{Anandtech}, ``{Comparing IPC on Skylake: Memory Latency and CPU Benchmarks},''
  online, accessed July 2021,
  \url{https://www.anandtech.com/show/9483/intel-skylake-review-6700k-6600k-ddr4-ddr3-ipc-6th-generation/9}.

\bibitem{energy_star}
E.~STAR, ``{Computers Specification Version 8.0},'' online, accessed Jul 2021
  \url{https://bit.ly/3x44nxu}.

\bibitem{energy2014energy}
S.~ENERGY, ``{ENERGY STAR{\textregistered} Program Requirements for Computers
  Partner Commitments},'' 2014.

\bibitem{intel_rmt}
{Intel}, ``{Intel Ready Mode Technology (Intel RMT)},'' accessed Jul 2021,
  \url{https://intel.ly/3ryAwMq}.

\bibitem{bolla2016assessing}
R.~Bolla \emph{et~al.}, ``{Assessing the Potential for Saving Energy by
  Impersonating Idle Networked Devices},'' \emph{IEEE Journal on Selected Areas
  in Communications}, 2016.

\bibitem{haj2016fine}
J.~Haj-Yihia \emph{et~al.}, ``{Fine-grain Power Breakdown of Modern
  Out-of-order Cores and its Implications on Skylake-based Systems},''
  \emph{TACO}, 2016.

\bibitem{11_fayneh20164}
E.~Fayneh \emph{et~al.}, ``{4.1 14nm 6th-{G}eneration Core Processor SoC with
  Low Power Consumption and Improved Performance},'' in \emph{{ISSCC}}, 2016.

\bibitem{yasin2019metric}
A.~Yasin \emph{et~al.}, ``{A Metric-Guided Method for Discovering Impactful
  Features and Architectural Insights for Skylake-Based Processors},''
  \emph{TACO}, 2019.

\bibitem{gonzalez1996energy}
R.~Gonzalez and M.~Horowitz, ``{Energy Dissipation in General Purpose
  Microprocessors},'' \emph{JSSC}, 1996.

\bibitem{hajsysscale}
J.~Haj-Yahya \emph{et~al.}, ``{SysScale: Exploiting Multi-{D}omain Dynamic
  Voltage and Frequency Scaling for Energy E{ffi}cient Mobile Processors},'' in
  \emph{ISCA}, 2020.

\bibitem{rotem2011power}
E.~Rotem \emph{et~al.}, ``{Power Management Architecture of the 2nd Generation
  Intel{\textregistered} Core Microarchitecture, Formerly Codenamed Sandy
  Bridge},'' in \emph{HotChips}, 2011.

\bibitem{rotem2012power}
E.~Rotem \emph{et~al.}, ``{Power-management Architecture of the Intel
  Microarchitecture Code-named Sandy Bridge},'' \emph{IEEE MICRO}, 2012.

\bibitem{rotem2015intel}
E.~Rotem, ``{Intel Architecture, Code Name Skylake Deep Dive: A New
  Architecture to Manage Power Performance and Energy Efficiency},'' in
  \emph{Intel Developer Forum}, 2015.

\bibitem{8th_9th_gen_intel}
{Intel}, ``{{8th and 9th Generation Intel Core Processor Families Datasheet,
  Volume 1 of 2}},'' online, accessed Aug 2020 \url{https://intel.ly/30VcShP}.

\bibitem{icelake2020}
{Intel}, ``{Ice Lake, 10th Generation Intel® Core™ Processor Families},''
  \url{https://intel.ly/3frvxpK"}, July 2019.

\bibitem{perf_limit_reasons}
{Intel}, ``{{6th Generation Intel Processor Families for S-Platforms}},''
  online accessed Aug 2020 \url{https://intel.ly/2XVdORo}.

\bibitem{haj2018energy}
J.~Haj-Yahya \emph{et~al.}, \emph{Energy Efficient High Performance Processors:
  Recent Approaches for Designing Green High Performance Computing}.\hskip 1em
  plus 0.5em minus 0.4em\relax Springer, 2018.

\bibitem{6_kanter2013haswell}
D.~Kanter, ``{Haswell FIVR Extends Battery Life},'' \emph{Microprocessor
  Report, The Linley Group}, 2013.

\bibitem{kahng2013many}
A.~B. Kahng \emph{et~al.}, ``{Many-Core Token-Based Adaptive Power Gating},''
  \emph{TCAD}, 2013.

\bibitem{chadha2013architectural}
R.~Chadha and J.~Bhasker, ``{Architectural Techniques for Low Power},'' in
  \emph{An ASIC Low Power Primer}.\hskip 1em plus 0.5em minus 0.4em\relax
  Springer, 2013.

\bibitem{usami2009design}
K.~Usami \emph{et~al.}, ``{Design and Implementation of Fine-grain Power Gating
  with Ground Bounce Suppression},'' in \emph{VLSI Design}, 2009.

\bibitem{agarwal2006power}
K.~Agarwal \emph{et~al.}, ``{Power Gating With Multiple Sleep Modes},'' in
  \emph{ISQED}, 2006.

\bibitem{abba2014improved}
A.~Abba and K.~Amarender, ``{Improved Power Gating Technique for Leakage Power
  Reduction},'' \emph{International Journal of Engineering and Science}, 2014.

\bibitem{larsson1997di}
P.~Larsson, ``{di/dt Noise in CMOS Integrated Circuits},'' in \emph{Analog
  Design Issues in Digital VLSI Circuits and Systems}.\hskip 1em plus 0.5em
  minus 0.4em\relax Springer, 1997.

\bibitem{akl2009effective}
C.~J. Akl \emph{et~al.}, ``{An Effective Staggered-Phase Damping Technique for
  Suppressing Power-Gating Resonance Noise During Mode Transition},'' in
  \emph{ISQED}, 2009.

\bibitem{kahng2012tap}
A.~B. Kahng \emph{et~al.}, ``{TAP: Token-based Adaptive Power Gating},'' in
  \emph{ISLPED}, 2012.

\bibitem{lempel20112nd}
O.~Lempel, ``{2nd Generation Intel{\textregistered} Core Processor Family:
  Intel{\textregistered} Core i7, i5 and i3},'' in \emph{Hot Chips}, 2011.

\bibitem{rotem2013power}
E.~Rotem \emph{et~al.}, ``{Power and Thermal Constraints of Modern
  System-on-a-Chip Computer},'' in \emph{THERMINIC}, 2013.

\bibitem{21_doweck2017inside}
J.~Doweck \emph{et~al.}, ``{Inside 6th-Generation Intel Core: New
  Microarchitecture Code-Named Skylake},'' \emph{IEEE Micro}, vol.~37, no.~2,
  pp. 52--62, 2017.

\bibitem{ranganathan2006ensemble}
P.~Ranganathan \emph{et~al.}, ``{Ensemble-level Power Management for Dense
  Blade Servers},'' \emph{ISCA}, 2006.

\bibitem{zhang2016maximizing}
H.~Zhang and H.~Hoffmann, ``{Maximizing Performance Under a Power Cap: A
  Comparison of Hardware, Software, and Hybrid Techniques},'' in \emph{ASPLOS},
  2016.

\bibitem{isci2006analysis}
C.~Isci \emph{et~al.}, ``{An Analysis of Efficient Multi-core Global Power
  Management Policies: Maximizing Performance for a Given Power Budget},'' in
  \emph{MICRO}, 2006.

\bibitem{ananthakrishnan2014dynamically}
A.~N. Ananthakrishnan \emph{et~al.}, ``{Dynamically allocating a power budget
  over multiple domains of a processor},'' Jul.~1 2014, uS Patent 8,769,316.

\bibitem{haj2020flexwatts}
J.~Haj-Yahya \emph{et~al.}, ``{FlexWatts: A Power-and Workload-Aware Hybrid
  Power Delivery Network for Energy-Efficient Microprocessors},'' in
  \emph{MICRO}, 2020.

\bibitem{kim2008system}
W.~Kim \emph{et~al.}, ``{System Level Analysis of Fast, per-core DVFS using
  On-Chip Switching Regulators},'' in \emph{2008 IEEE 14th International
  Symposium on High Performance Computer Architecture (ISCA)}.\hskip 1em plus
  0.5em minus 0.4em\relax IEEE, 2008, pp. 123--134.

\bibitem{acpi}
{UEFI.org}, ``{{Advanced Configuration and Power Interface (ACPI) specification
  }},'' online, accessed July 2021, \url{https://bit.ly/2ToagrG}.

\bibitem{amd_specs}
{AMD}, ``{AMD Processor Specifications},'' online, accessed June 2020,
  https://www.amd.com/en/products/specifications/processors.

\bibitem{qcom2018}
{Qualcomm Technologies}, ``{{Qualcomm Snapdragon 410E (APQ8016E) Processor
  Device Specification}},'' online, 2018,
  https://developer.qualcomm.com/qfile/28813/lm80-p0436-7\_f\_410e\_proc\_apq8016e\_device\_spec.pdf.

\bibitem{psr}
S.~Kwa \emph{et~al.}, ``{Panel Self-Refresh Technology: Decoupling Image Update
  from LCD Panel Refresh in Mobile Computing Systems},'' in \emph{SID Symposium
  Digest of Technical Papers}, 2012.

\bibitem{haj-micro-2021}
J.~Haj-Yahya \emph{et~al.}, ``{BurstLink: Techniques for Energy-Efficient Video
  Display for Conventional and Virtual Reality Systems},'' in \emph{MICRO},
  2021.

\bibitem{hammarlund2014haswell}
P.~Hammarlund \emph{et~al.}, ``{Haswell: The Fourth-{G}eneration Intel Core
  Processor},'' \emph{IEEE Micro}, 2014.

\bibitem{mosalikanti2015low}
P.~Mosalikanti \emph{et~al.}, ``{Low Power Analog Circuit Techniques in the 5th
  Generation Intel Core Microprocessor (Broadwell)},'' in \emph{CICC}, 2015.

\bibitem{kurd2014haswell}
N.~Kurd \emph{et~al.}, ``{Haswell: A Family of IA 22 nm Processors},''
  \emph{JSSC}, 2014.

\bibitem{deval2015power}
A.~Deval \emph{et~al.}, ``{Power Management on 14 nm Intel{\textregistered}
  Core- M processor},'' in \emph{COOL CHIPS}, 2015.

\bibitem{Skylake_die_server}
{Wikichip}, ``{Skylake (server) - Microarchitectures - Intel},'' online,
  accessed {August 2019 \url{https://bit.ly/2MHEWkj}}.

\bibitem{Skylake_H}
{Intel}, ``{{6th Generation Intel® Core™ Processor for H-Platforms}},''
  online accessed Jul 2021 \url{https://intel.ly/2XVdORo}.

\bibitem{skl_dies}
{Intel}, ``{{Intel Skylake client - Microarchitectures - Dies}},'' online
  accessed Jul 2021 \url{https://bit.ly/3kPSdpK}.

\bibitem{tam2018skylake}
S.~M. Tam \emph{et~al.}, ``{Skylake-SP: A 14nm 28-Core Xeon{\textregistered}
  Processor},'' in \emph{ISSCC}, 2018.

\bibitem{intel_skl__3_5}
{Intel}, ``{Intel Core m5-6Y57 Processor},'' online, accessed Nov 2019,
  https://ark.intel.com/content/www/us/en/ark/products/88197/intel-core-m5-6y57-processor-4m-cache-up-to-2-80-ghz.html.

\bibitem{intel_skl_91}
{Intel}, ``{Intel® Core™ i7-6700K Processor},'' online, accessed Nov 2019,
  https://intel.ly/36w8d7U.

\bibitem{singh20173}
T.~Singh \emph{et~al.}, ``{3.2 Zen: A Next-generation High-performance
  $\times$86 Core},'' in \emph{ISSCC}, 2017.

\bibitem{singh2018zen}
T.~Singh \emph{et~al.}, ``{Zen: An Energy-Efficient High-Performance $-x86$
  Core},'' \emph{JSSC}, 2018.

\bibitem{burd2019zeppelin}
T.~Burd \emph{et~al.}, ``{Zeppelin: An SoC for Multichip Architectures},''
  \emph{JSSC}, 2019.

\bibitem{beck2018zeppelin}
N.~Beck \emph{et~al.}, ``{Zeppelin: An SoC for Multichip Architectures},'' in
  \emph{{ISSCC}}, 2018.

\bibitem{amd_zen2_10}
{AMD}, ``{AMD Ryzen 3 4300U},'' online, accessed April 2020,
  https://www.amd.com/en/products/apu/amd-ryzen-3-4300u.

\bibitem{amd_zen2_54}
{AMD}, ``{AMD Ryzen 7 4800H},'' online, accessed April 2020,
  https://www.amd.com/en/products/apu/amd-ryzen-7-4800H.

\bibitem{cTDP}
{Wikipedia}, ``{Configurable TDP},'' online, accessed March 2018, Mar 2019,
  https://en.wikipedia.org/wiki/Thermal\_design\_power.

\bibitem{cTDP2}
{Anandtech}, ``{Configurable TDP},'' online, accessed March 2020,
  https://www.anandtech.com/show/4830/intels-ivy-bridge-architecture-exposed/4.

\bibitem{jahagirdar2012power}
S.~Jahagirdar \emph{et~al.}, ``{Power Management of the Third Generation Intel
  Core Micro Architecture Formerly Codenamed Ivy Bridge},'' in \emph{HotChips},
  2012.

\bibitem{kujala2002transition}
A.~Kujala \emph{et~al.}, ``{Transition to Pb-free Manufacturing Using Land Grid
  Array Packaging Technology},'' in \emph{ECTC}, 2002.

\bibitem{guenin1995analysis}
B.~M. Guenin \emph{et~al.}, ``{Analysis of a Thermally Enhanced Ball Grid Array
  Package},'' \emph{CPMT}, 1995.

\bibitem{magarshack2003system}
P.~Magarshack and P.~G. Paulin, ``{System-on-chip Beyond the Nanometer Wall},''
  in \emph{DAC}, 2003.

\bibitem{haj2021ichannels}
J.~Haj-Yahya \emph{et~al.}, ``{IChannels: Exploiting Current Management
  Mechanisms to Create Covert Channels in Modern Processors},'' \emph{ISCA},
  2021.

\bibitem{10_jahagirdar2012power}
S.~Jahagirdar \emph{et~al.}, ``{Power Management of the Third Generation Intel
  Core Micro Architecture Formerly Codenamed Ivy Bridge},'' in \emph{HotChips},
  2012.

\bibitem{12_howse2015tick}
B.~Howse and R.~Smith, ``{Tick Tock On The Rocks: Intel Delays 10nm, Adds 3rd
  Gen 14nm Core Product Kaby Lake},'' 2015.

\bibitem{toprak20145}
Z.~Toprak-Deniz \emph{et~al.}, ``{5.2 Distributed System of Digitally
  Controlled Microregulators Enabling per-core DVFS for the POWER8 TM
  Microprocessor},'' in \emph{ISSCC}, 2014.

\bibitem{sinkar2013low}
A.~A. Sinkar \emph{et~al.}, ``{Low-cost Per-core Voltage Domain Support for
  Power-constrained High-{P}erformance Processors},'' \emph{VLSI}, 2013.

\bibitem{5_nalamalpu2015broadwell}
A.~Nalamalpu \emph{et~al.}, ``{Broadwell: A Family of IA 14nm Processors},'' in
  \emph{VLSI Circuits}, 2015.

\bibitem{mandelblat2015technology}
J.~Mandelblat, ``{Technology Insight: Intel’s Next Generation
  Microarchitecture Code Name Skylake},'' in \emph{Intel Developer Forum, San
  Francisco}, 2015.

\bibitem{intel_avx512}
J.~Reinders, ``{Intel AVX-512 Instructions},'' \emph{Intel Software Developer
  Zone, Jun}, 2017.

\bibitem{14_module2009and}
Intel, ``{Module, Voltage Regulator and Enterprise Voltage Regulator-Down
  (EVRD) 11.1 Design Guidelines},'' \emph{Intel Corp., Santa Clara, CA}, 2009.

\bibitem{intel_avp_2009}
{Intel}, ``{Voltage Regulator Module (VRM) and Enterprise Voltage
  Regulator-Down (EVRD) 11.1}.''

\bibitem{sun2006novel}
J.~Sun \emph{et~al.}, ``{A Novel Input-side Current Sensing Method to Achieve
  AVP for Future VRs},'' \emph{IEEE Transactions on Power Electronics}, 2006.

\bibitem{tsai2015switching}
C.-H. Tsai \emph{et~al.}, ``{Switching Frequency Stabilization Techniques for
  Adaptive on-time Controlled Buck Converter with Adaptive Voltage Positioning
  Mechanism},'' \emph{IEEE T{PEL}}, 2015.

\bibitem{reddi2009voltage}
V.~J. Reddi \emph{et~al.}, ``{Voltage Emergency Prediction: Using Signatures to
  Reduce Operating Margins},'' in \emph{HPCA}, 2009.

\bibitem{reddi2010voltage}
V.~J. Reddi \emph{et~al.}, ``{Voltage Smoothing: Characterizing and Mitigating
  Voltage Noise in Production Processors via Software-guided Thread
  Scheduling},'' in \emph{MICRO}, 2010.

\bibitem{peterchev2006load}
A.~V. Peterchev and S.~R. Sanders, ``{Load-line Regulation with Estimated
  Load-current Feedforward: Application to Microprocessor Voltage
  Regulators},'' \emph{IEEE T{PEL}}, 2006.

\bibitem{haj2015compiler}
J.~Haj-Yihia \emph{et~al.}, ``{Compiler-directed Power Management for
  Superscalars},'' \emph{TACO}, 2015.

\bibitem{fetzer2015managing}
E.~Fetzer \emph{et~al.}, ``{Managing Power Consumption in a Multi-core
  Processor},'' 2015, {US Patent 9,069,555}.

\bibitem{intel_xtu_overclocking}
{Intel}, ``{{Overclocking Intel® Core Processors: Taking Overclocking to the
  Next Level}},'' online, accessed Aug 2020, https://bit.ly/3iITafa.

\bibitem{xie2014therminator}
Q.~Xie \emph{et~al.}, ``{Therminator: A Thermal Simulator for Smartphones
  Producing Accurate Chip and Skin Temperature Maps},'' in \emph{ISLPED}, 2014.

\bibitem{rotem2015power}
E.~Rotem \emph{et~al.}, ``{Power and Thermal Constraints of Modern
  System-on-a-Chip Computer},'' \emph{Microelectronics Journal}, 2015.

\bibitem{su2017high}
Y.~Su \emph{et~al.}, ``{High-efficiency Multiphase DC-DC Converters for
  Powering Processors with Turbo Mode Based on Configurable Current Sharing
  Ratios and Intelligent Phase Management},'' in \emph{APEC}, 2017.

\bibitem{meisner2009powernap}
D.~Meisner \emph{et~al.}, ``{Powernap: Eliminating Server Idle Power},''
  \emph{ASPLOS}, 2009.

\bibitem{skylakex}
{Intel}, ``{Skylake-X, 6th Generation Intel Core X-series Processors
  Families},'' \url{https://intel.ly/30SP8uX}, July 2019.

\bibitem{naffziger2016integrated}
S.~Naffziger, ``{Integrated Power Conversion Strategies Across Laptop Server
  and Graphics Products},'' in \emph{Proc. Power Supply Chip}, 2016,
  \url{https://bit.ly/3rEbhIR}.

\bibitem{wright2006characterization}
S.~Wright \emph{et~al.}, ``{Characterization of Micro-bump C4 Interconnects for
  Si-carrier SOP Applications},'' in \emph{ECTC}, 2006.

\bibitem{piguet2005low}
C.~Piguet, \emph{{Low-Power CMOS Circuits: Technology, Logic Design and CAD
  Tools}}.\hskip 1em plus 0.5em minus 0.4em\relax CRC Press, 2005.

\bibitem{zhang2014architecture}
R.~Zhang \emph{et~al.}, ``{Architecture Implications of Pads as a Scarce
  Resource},'' in \emph{ISCA}, 2014.

\bibitem{joseph2003control}
R.~Joseph \emph{et~al.}, ``{Control Techniques to Eliminate Voltage Emergencies
  in High Performance Processors},'' in \emph{HPCA}, 2003.

\bibitem{brooks2001dynamic}
D.~Brooks and M.~Martonosi, ``{Dynamic {T}hermal {M}anagement for
  {H}igh-{P}erformance {M}icroprocessors},'' in \emph{HPCA}, 2001.

\bibitem{lefurgy2011active}
C.~R. Lefurgy \emph{et~al.}, ``{Active Management of Timing Guardband to Save
  Energy in {POWER7}},'' in \emph{MICRO}, 2011.

\bibitem{hu1996gate}
C.~Hu, ``{Gate Oxide Scaling Limits and Projection},'' in \emph{IEDM}, 1996.

\bibitem{mercati2013workload}
P.~Mercati \emph{et~al.}, ``{Workload and User Experience-aware Dynamic
  Reliability Management in Multicore Processors},'' in \emph{DAC}, 2013.

\bibitem{intel_bios_overclocking}
{Intel}, ``{{How to Overclock Your CPU from BIOS}},'' online, accessed Aug 2020
  \url{https://www.intel.com/content/www/us/en/gaming/resources/bios-overclocking.html}.

\bibitem{SPEC2018}
{Standard Performance Evaluation Corporation}, ``{SPEC},'' online, accessed Jul
  2021, www.spec.org.

\bibitem{mansfeld1981recording}
F.~Mansfeld, ``{Recording and Analysis of AC Impedance Data for Corrosion
  Studies},'' \emph{Corrosion}, 1981.

\bibitem{leng2014gpuvolt}
J.~Leng \emph{et~al.}, ``{GPUVolt: Modeling and Characterizing Voltage Noise in
  GPU Architectures},'' in \emph{ISLPED}, 2014.

\bibitem{ketkar2009microarchitecture}
M.~Ketkar and E.~Chiprout, ``{A Microarchitecture-based Framework for Pre- and
  Post-silicon Power Delivery Analysis},'' in \emph{MICRO}, 2009.

\bibitem{zu2015adaptive}
Y.~Zu \emph{et~al.}, ``{Adaptive Guardband Scheduling to Improve System-Level
  Efficiency of the POWER7+},'' in \emph{MICRO}, 2015.

\bibitem{zou2018efficient}
A.~Zou \emph{et~al.}, ``{Efficient and Reliable Power Delivery in
  Voltage-stacked Manycore System with Hybrid Charge-recycling Regulators},''
  in \emph{DAC}, 2018.

\bibitem{leng2015gpu}
J.~Leng \emph{et~al.}, ``{GPU voltage noise: Characterization and Hierarchical
  Smoothing of Spatial and Temporal Voltage Noise Interference in GPU
  Architectures},'' in \emph{HPCA}, 2015.

\bibitem{song2014architectural}
W.~Song \emph{et~al.}, ``{Architectural reliability: Lifetime Reliability
  Characterization and Management of Many-core Processors},'' \emph{CAL}, 2014.

\bibitem{swaminathan2017bravo}
K.~Swaminathan \emph{et~al.}, ``{Bravo: Balanced Reliability-Aware Voltage
  Optimization},'' in \emph{HPCA}, 2017.

\bibitem{kim2016learning}
T.~Kim \emph{et~al.}, ``{Learning-based Dynamic Reliability Management for Dark
  Silicon Processor Considering EM Effects},'' in \emph{DATE}, 2016.

\bibitem{19_MSFT}
{Anandtech}, ``{The Microsoft Surface Pro (2017) Review: Evaluation},'' online,
  accessed March 2018, Mar 2019,
  https://www.anandtech.com/show/11538/the-microsoft-surface-pro-2017-review-evolution/7.

\bibitem{haj2020techniques}
J.~Haj-Yahya \emph{et~al.}, ``{Techniques for Reducing the Connected-Standby
  Energy Consumption of Mobile Devices},'' in \emph{HPCA}, 2020.

\bibitem{kulkarni2009high}
S.~H. Kulkarni \emph{et~al.}, ``{High-Density 3-D Metal-Fuse PROM Featuring
  1.37 $\mu$m 2 1T1R Bit Cell in 32nm High-k Metal-Gate CMOS Technology},'' in
  \emph{Symposium on VLSI Circuits}, 2009.

\bibitem{aygun2005power}
K.~Ayg{\"u}n \emph{et~al.}, ``{Power Delivery for High-Performance
  Microprocessors.}'' \emph{Intel Technology Journal}, 2005.

\bibitem{shekhar2016power}
S.~Shekhar \emph{et~al.}, ``{Power Delivery Impedance Impact of Power Gating
  Schemes},'' in \emph{SPI}, 2016.

\bibitem{i7_6700K}
{Intel}, ``{{Intel Core i7-6700K Processor}},'' online, accessed June 2021,
  https://intel.ly/3y4SWXA.

\bibitem{i7_6920HQ}
{Intel}, ``{{Intel Core i7-6920HQ Processor}},'' online, accessed June 2021,
  https://intel.ly/3xXrrzj.

\bibitem{ddr4_ref}
JEDEC, ``{DDR4 SDRAM Standard},'' \emph{JEDEC Std., JESD79-4C}, 2020.

\bibitem{intel_itp}
{Intel}, ``{{In Target Probe (ITP)}},'' online, accessed July 2021,
  https://bit.ly/2qHSMbm.

\bibitem{williams2009low}
M.~Williams, ``{Low Pin-Count Debug Interfaces for Multi-Device Systems},''
  \emph{ARM’s Serial Wire Debug white paper}, 2009.

\bibitem{itp_2020}
{Intel}, ``{Intel In-Target Probe - Extended Debug Port (Intel ITP-XDP)},''
  online, accessed December 2021, 2020, https://intel.ly/32c8DRI.

\bibitem{itp_xlsoft}
{Intel}, ``{Intel JTAG Debugger Quickstart Guide},'' 2013,
  https://www.xlsoft.com/jp/products/intel/system/2013/xdb-quickstart-lin.pdf.

\bibitem{NIDAQ}
{National Instruments}, ``{{NI-DAQ PCIe-6376}},'' online accessed 2019
  \url{http://www.ni.com/pdf/manuals/377387c.pdf}.

\bibitem{17_3DMARK}
{Vantage}, ``{3DMARK},'' online, accessed March 2018, Mar 2018,
  http://www.futuremark.com/benchmarks/3dmarkvantage.

\bibitem{intel_rmt2}
{Intel}, ``{Intel® Ready Mode Technology (Intel RMT)},'' online, accessed
  December 2021, https://intel.ly/3GQ7Jtb.

\bibitem{gough2015energy}
C.~Gough \emph{et~al.}, \emph{{Energy Efficient Servers: Blueprints for Data
  Center Optimization}}.\hskip 1em plus 0.5em minus 0.4em\relax Apress, 2015.

\bibitem{tu2015atom}
S.~Tu, ``Atom™-x5/x7 series processor, codenamed cherry trail,'' in
  \emph{2015 IEEE Hot Chips 27 Symposium (HCS)}.\hskip 1em plus 0.5em minus
  0.4em\relax IEEE, 2015, pp. 1--28.

\bibitem{haj2012transferring}
J.~Haj-Yihia, ``{Transferring Architectural Functions of a Processor to a
  Platform Control Hub Responsive to the Processor Entering a Deep Sleep
  State},'' Jul.~24 2012, uS Patent 8,230,247.

\bibitem{haj2013connected}
J.~Haj-Yihia, ``{Connected Standby Sleep State},'' Jun.~4 2013, uS Patent
  8,458,503.

\bibitem{rmt_anandtech}
{Anandtech}, ``{Intel Announces Ready Mode Technology: Using C7 for Syncing and
  Streaming},'' online, accessed December 2021,
  \url{https://www.anandtech.com/show/7871/intel-ready-mode-technology}.

\bibitem{yasin2017performance}
A.~Yasin \emph{et~al.}, ``{Performance Scalability Prediction},'' Nov.~28 2017,
  {US Patent 9,829,957}.

\bibitem{haj2015doee}
J.~Haj-Yihia \emph{et~al.}, ``{DOEE: Dynamic Optimization Framework for Better
  Energy Efficiency},'' in \emph{HiPC}, 2015.

\bibitem{kim2008silicon}
T.-H. Kim \emph{et~al.}, ``{Silicon Odometer: An On-chip Reliability Monitor
  for Measuring Frequency Degradation of Digital Circuits},'' \emph{JSSC},
  2008.

\bibitem{cho2015aging}
M.~Cho \emph{et~al.}, ``{Aging-aware Adaptive Voltage Scaling in 22nm
  high-K/metal-gate tri-gate CMOS},'' in \emph{CICC}, 2015.

\bibitem{leng2015safe}
J.~Leng \emph{et~al.}, ``{Safe Limits on Voltage Reduction Efficiency in GPUs:
  a Direct Measurement Approach},'' in \emph{MICRO}, 2015.

\bibitem{bacha2013dynamic}
A.~Bacha and R.~Teodorescu, ``{Dynamic Reduction of Voltage Margins by
  Leveraging On-chip ECC in Itanium II Processors},'' in \emph{ISCA}, 2013.

\bibitem{von1990voltage}
V.~Von~Kaenel \emph{et~al.}, ``{A Voltage Reduction Technique for
  Battery-operated Systems},'' \emph{IEEE Journal of Solid-State Circuits},
  1990.

\bibitem{lefurgy2013active}
C.~R. Lefurgy \emph{et~al.}, ``{Active Guardband Management in Power7+ to Save
  Energy and Maintain Reliability},'' \emph{IEEE Micro}, 2013.

\bibitem{bertran2014voltage}
R.~Bertran \emph{et~al.}, ``{Voltage Noise in Multi-core Processors: Empirical
  Characterization and Optimization Opportunities},'' in \emph{MICRO}, 2014.

\bibitem{papadimitriou2017harnessing}
G.~Papadimitriou \emph{et~al.}, ``{Harnessing Voltage Margins for Energy
  Efficiency in Multicore CPUs},'' in \emph{MICRO}, 2017.

\bibitem{thomas2016core}
R.~Thomas \emph{et~al.}, ``{Core tunneling: Variation-aware voltage noise
  mitigation in GPUs},'' in \emph{2016 IEEE International Symposium on High
  Performance Computer Architecture (HPCA)}.\hskip 1em plus 0.5em minus
  0.4em\relax IEEE, 2016, pp. 151--162.

\bibitem{miller2012vrsync}
T.~N. Miller \emph{et~al.}, ``{VRSync: Characterizing and Eliminating
  Synchronization-induced Voltage Emergencies in Many-core Processors},'' in
  \emph{ISCA}, 2012.

\bibitem{leng2020predictive}
J.~Leng \emph{et~al.}, ``{Predictive Guardbanding: Program-driven Timing Margin
  Reduction for GPUs},'' \emph{TCAD}, 2020.

\bibitem{ernst2003razor}
D.~Ernst \emph{et~al.}, ``{Razor: A Low-power Pipeline Based on Circuit-level
  Timing Speculation},'' in \emph{MICRO}, 2003.

\bibitem{powell2003pipeline}
M.~D. Powell and T.~Vijaykumar, ``{Pipeline Damping: a Microarchitectural
  Technique to Reduce Inductive Noise in Supply Voltage},'' in \emph{ISCA},
  2003.

\bibitem{gupta2007understanding}
M.~S. Gupta \emph{et~al.}, ``{Understanding Voltage Variations in Chip
  Multiprocessors using a Distributed Power-delivery Network},'' in
  \emph{DATE}, 2007.

\bibitem{engin2010efficient}
A.~E. Engin, ``{Efficient Sensitivity Calculations for Optimization of Power
  Delivery Network Impedance},'' \emph{TEMC}, 2010.

\bibitem{das2008razorii}
S.~Das \emph{et~al.}, ``{RazorII: In Situ Error Detection and Correction for
  PVT and SER Tolerance},'' \emph{JSSC}, 2008.

\bibitem{das2006self}
S.~Das \emph{et~al.}, ``{A Self-tuning DVS Processor Using Delay-error
  Detection and Correction},'' \emph{JSSC}, 2006.

\bibitem{kim2012audit}
Y.~Kim \emph{et~al.}, ``{AUDIT: Stress Testing the Automatic Way},'' in
  \emph{MICRO}, 2012.

\end{thebibliography}
%%%%%%%%%%%%%%%%%%%%%%%%%%%%%%%%%%%%

% that's all folks
\end{document}